\documentclass[12pt,tightenlines,nofootinbib,aps,floatfix,pra,superscriptaddress,eqsecnum]{revtex4-1}

\usepackage{booktabs}
\usepackage{dcolumn}
\usepackage{units}
\usepackage{array}

\usepackage[utf8]{inputenc}
\usepackage{amsfonts}
\usepackage{amsmath,amssymb,bm}
\usepackage{graphicx}
\usepackage{mathtools}
\usepackage{bbold}
\usepackage[section]{placeins}
\usepackage{subcaption}
\usepackage{ragged2e}
\DeclareCaptionJustification{justified}{\justifying}
\usepackage{alphalph}
\usepackage{natbib}
\usepackage{textcomp}
\usepackage{braket}
\usepackage{slashed}
\usepackage{color}
\usepackage{float}
\usepackage{subfig}
\usepackage{multirow}
\usepackage{hhline}
\usepackage{subfig}

\def\){\right)} 
\def\({\left(} 
\def\]{\right]} 
\def\[{\left[}

\newcommand{\spinup}{\uparrow}
\newcommand{\spindown}{\downarrow}



\begin{document}

\title{Nucleon-deuteron scattering \\ using the adiabatic projection method}

\author{Serdar Elhatisari}
\email{elhatisari@hiskp.uni-bonn.de}
\affiliation{Helmholtz-Institut f\"ur Strahlen- und Kernphysik~(Theorie) 
and Bethe Center for Theoretical Physics, Universit\"at Bonn, D-53115 Bonn, Germany}

\author{Dean Lee}
\email{dean\_lee@ncsu.edu}
\affiliation{Department of Physics, North Carolina State University,
Raleigh, North Carolina 27695, USA}

\author{Ulf-G. Mei{\ss}ner}
\email{meissner@hiskp.uni-bonn.de}
\affiliation{Helmholtz-Institut f\"ur Strahlen- und Kernphysik~(Theorie) 
	and Bethe Center for Theoretical Physics, Universit\"at Bonn, D-53115 Bonn, Germany}
\affiliation{Institute for Advanced Simulation, Institut f{\"u}r Kernphysik, and J{\"u}lich Center for Hadron Physics, Forschungszentrum J{\"u}lich, D-52425 J{\"u}lich, Germany}
\affiliation{JARA - High Performance Computing, Forschungszentrum J\"ulich, D-52425 J\"ulich, Germany}

\author{Gautam Rupak}
\email{grupak@u.washington.edu}
\affiliation{Department of Physics $\&$ Astronomy and 
HPC$^2$ Center for Computational Sciences, 
Mississippi State
University, Mississippi State, Mississippi State 39762, USA}

\date{\today}

\begin{abstract}
In this paper we discuss the adiabatic projection method, a general framework for scattering and reaction calculations on the lattice. We also introduce several new techniques developed to study nucleus-nucleus scattering and reactions on the lattice. We present technical details of the method for large-scale problems. To estimate the systematic errors of the calculations we consider simple two-particle scattering on the lattice. Then we  benchmark the accuracy and efficiency of the numerical methods by applying these to calculate fermion-dimer scattering in lattice effective field theory with and without a long-range Coulomb potential. The fermion-dimer calculations correspond to neutron-deuteron and proton-deuteron scattering in the spin-quartet channel at leading order in the pionless effective field theory. 
\end{abstract}

\maketitle

%=============================  Introduction ==========================
\section{Introduction}
\label{sec:introduction}
%=================================================================

One of the major challenges in computational strong interaction physics is the \emph{ab initio} calculation of nucleus-nucleus scattering and reactions. In the last decade there has been significant progress on nucleon-nucleus~\cite{Nollett:2006su,Quaglioni:2008sm,Navratil:2010jn,Navratil:2011sa,Hupin:2013wsa,Hagen:2012rq} and nucleus-nucleus~\cite{Navratil:2011ay,Navratil:2011zs,Elhatisari:2015iga} scattering and reactions. Recently the scattering of two ${}^{4}$He (alpha) clusters has been achieved on the lattice using the adiabatic projection method, and so far the ${}^{4}$He nuclei has been the heaviest projectile used in the \emph{ab initio} calculations of scattering and reactions~\cite{Elhatisari:2015iga}. This was an important step towards making the \emph{ab initio} description of scattering and reactions involving heavier nuclei practical. It also opens the doors towards using experimental data from collisions of heavier nuclei as input to improve \emph{ab initio} nuclear structure theory. See Ref.~\cite{Elhatisari:2016owd} for a recent study that uses the ${}^{4}$He~-${}^{4}$He scattering as a tool for probing the nuclear structure of alpha-like nuclei which are nuclei with equal and even numbers of protons and neutrons.

The adiabatic projection method is a general framework for scattering and reaction calculations on the lattice, and the method splits the calculation into two parts~\cite{Pine:2013zja,Elhatisari:2014lka,Rokash:2015hra}. In the first part, the method starts from the microscopic Hamiltonian and uses initial cluster states and Euclidean time projection to construct a low-energy effective theory for the clusters, the so-called adiabatic Hamiltonian. The description of the low-lying two-cluster states from the adiabatic Hamiltonian becomes exact in the limit of large Euclidean projection time. In the second part of the method, the adiabatic Hamiltonian that describes the low-energy interaction is used to calculate elastic and inelastic reactions such as $a(b,b)a$, $a(b,c)d$ and $a(b,\gamma)c$, where $a$, $b$, $c$ are nuclear clusters and $\gamma$ is a photon.  Elastic neutron-deuteron scattering $d(n,n)d$ has been calculated accurately in Refs.~\cite{Pine:2013zja,Elhatisari:2014lka,Rokash:2015hra}. The adiabatic projection method in the presence of long range Coulomb in alpha-like nuclei was considered in Refs.~\cite{Elhatisari:2015iga,Elhatisari:2016owd} that would allow future studies involving nuclear clusters that necessarily involve the Coulomb repulsion. The lattice method for calculating nuclear reactions  from an effective two-cluster Hamiltonian with an external electro-weak current in radiative neutron capture  $p(n,\gamma)d$ and proton-proton fusion $p(p,e^+\nu_e)d$ were also considered~\cite{Rupak:2013aue,Rupak:2014xza}.

In this paper, we present technical details of the adiabatic projection method and introduce several new techniques that improve the accuracy and efficiency of the method. We also discuss the systematic errors in the scattering and reaction calculations arising from the techniques developed here.

We organize our paper as follows. We start with the continuum and lattice formulations in Section~\ref{sec:Hamiltonian}. Then, in Section~\ref{sec:adiabatic-projection} we discuss the adiabatic projection formulation and describe the implementation of several new methods useful for adiabatic projection calculations. Section~\ref{sec:Scattering-phase-shifts} is dedicated to the discussion of the method used to extract the phase shifts from the lattice. In Section~\ref{appndx:two-particle-scattering} we implement the adiabatic projection formalism using new improved methods for two-particle scattering on the lattice and study the systematic errors of the calculations. Then in Section~\ref{sec:fermion-dimer-scattering} we consider fermion-dimer scattering to benchmark the adiabatic projection formalism against continuum calculations for fermion-dimer scattering.  The dimer is a two-particle bound state in these calculations. Finally, the results of our paper are summarized in Section~\ref{sec:summary-and-discussion}.

%========================== Lattice Hamiltonian ============================
\section{Lattice Hamiltonian}
\label{sec:Hamiltonian}
%======================================================================
The free nonrelativistic Hamiltonian in the continuum is,
%------------ Equation -------------------------
\begin{align}
\hat{H}_{0} 
= 
\sum_{s}
\frac{1}{2m_{s}}
\int d^{3}\vec{r} \, 
\vec{\nabla}b^{\dagger}_{s}(\vec{r})
\cdot
\vec{\nabla}b^{\,}_{s}(\vec{r})\,,
\label{eqn:free-Hamiltonian-001}
\end{align} 
where $s$ labels the particle species, and $b^{\,}_{s}$ and $b_{s}^{\dagger}$ are the annihilation and creation operators, respectively. 
The two-particle potential is written as
%------------ Equation -------------------------
\begin{align}
\hat{V} 
= \frac{1}{2} \sum_{s,s\,'} \, \int \int d^{3}\vec{r}\,' \, d^{3}\vec{r} \,
\rho_{s}(\vec{r}\,') \, V_{ss'}(\vec{r}-\vec{r}\,')\,\rho_{s'}(\vec{r})
\,,
\label{eqn:Interaction-001}
\end{align} 
with $\rho_{s}(\vec{r})$ the density operator,
%------------ Equation -------------------------
\begin{align}
\rho_{s}(\vec{r})
= b^{\dagger}_{s}(\vec{r}) b^{\,}_{s}(\vec{r})\,.
\label{eqn:density-operator-001}
\end{align}
In the lattice calculation, we take the particles to be spin-$\frac{1}{2}$ fermions that form a $s$-wave spin-singlet bound state identified as the dimer. At low-energy three-particle and higher-body potentials are not needed. The low-energy physics can be described in the effective field theory (EFT) with just a short-ranged interaction between the fermions at leading order, the so called pionless EFT~\cite{Bedaque:1997qi,Chen:1999tn}. In this theory the spin indices can be identified with the isospin indices of the nucleon-deuteron system in the spin quartet channel that we study, as we explain further below. We study systems consisting of nucleons on the lattice by discretizing space and time. We set the spatial lattice spacing to $a =$ 1.97 fm and the temporal lattice spacing is $a_{t} =$ 1.32 fm. We define all physical quantities in lattice units (l.u.) multiplying the quantities by the corresponding powers of $a$. Also, our lattice is periodic in all spatial directions. The free nonrelativistic lattice Hamiltonian with the $\mathcal{O}(a^{4})$-improved action is 
 %------------ Equation -------------------------
\begin{align}
\hat{H}_{0} 
= \sum_{s}
\,
\frac{1}{2m_{s}}
\,
\sum_{\hat{l}=\hat{x},\hat{y},\hat{z}}
\,
\sum_{\vec{n}}
\, 
\[
\sum_{k=-3}^{3}\, w_{|k|} \, 
b^{\dagger}_{s}(\vec{n})b^{\,}_{s}(\vec{n}+k\,\hat{l})
\]\,,
\label{eqn:free-Hamiltonian-002}
\end{align} 
where $w_{0}$, $w_{1}$, $w_{2}$, and $w_{3}$ are 49/18, -3/2, 3/20, and -1/90, respectively. $\vec{n}$ denotes the lattice sites, and the discretized form of the potential is
%------------ Equation -------------------------
\begin{align}
\hat{V} 
= \frac{1}{2} \sum_{s,s'}
\sum_{\vec{n}\,'}
\sum_{\vec{n}}
\rho_{s}(\vec{n}\,')
\,
V_{ss'}({\vec{n}\,'}-{\vec{n}})
\,
\rho_{s'}(\vec{n})
\,.
\label{eqn:Interaction-002}
\end{align} 

Since we discretize the temporal direction, we use the transfer matrix formalism and define normal-ordered transfer matrix operator $\hat{M}$ as,
%------------ Equation -------------------------
\begin{align}
\hat{M}
=
:
\exp\[- \alpha_{t}\, (\hat{H}_{0} -\hat{V}) \]
:\,,
\label{eqn:Transfer-matrix-001}
\end{align}
where $\alpha_{t} = a_{t}/a$, and the symbol $:\,:$ signifies the normal ordering of operators with annihilation operators on the right side and creation operators on the left side. The transfer matrix defined here is called the microscopic transfer matrix as it describes the full microscopic details of the system.

%=====================  Adiabatic Projection Method ======================
\section{Adiabatic projection method}
\label{sec:adiabatic-projection}
%==================================================

\begin{figure}[!ht]
\includegraphics[width=0.4\textwidth]{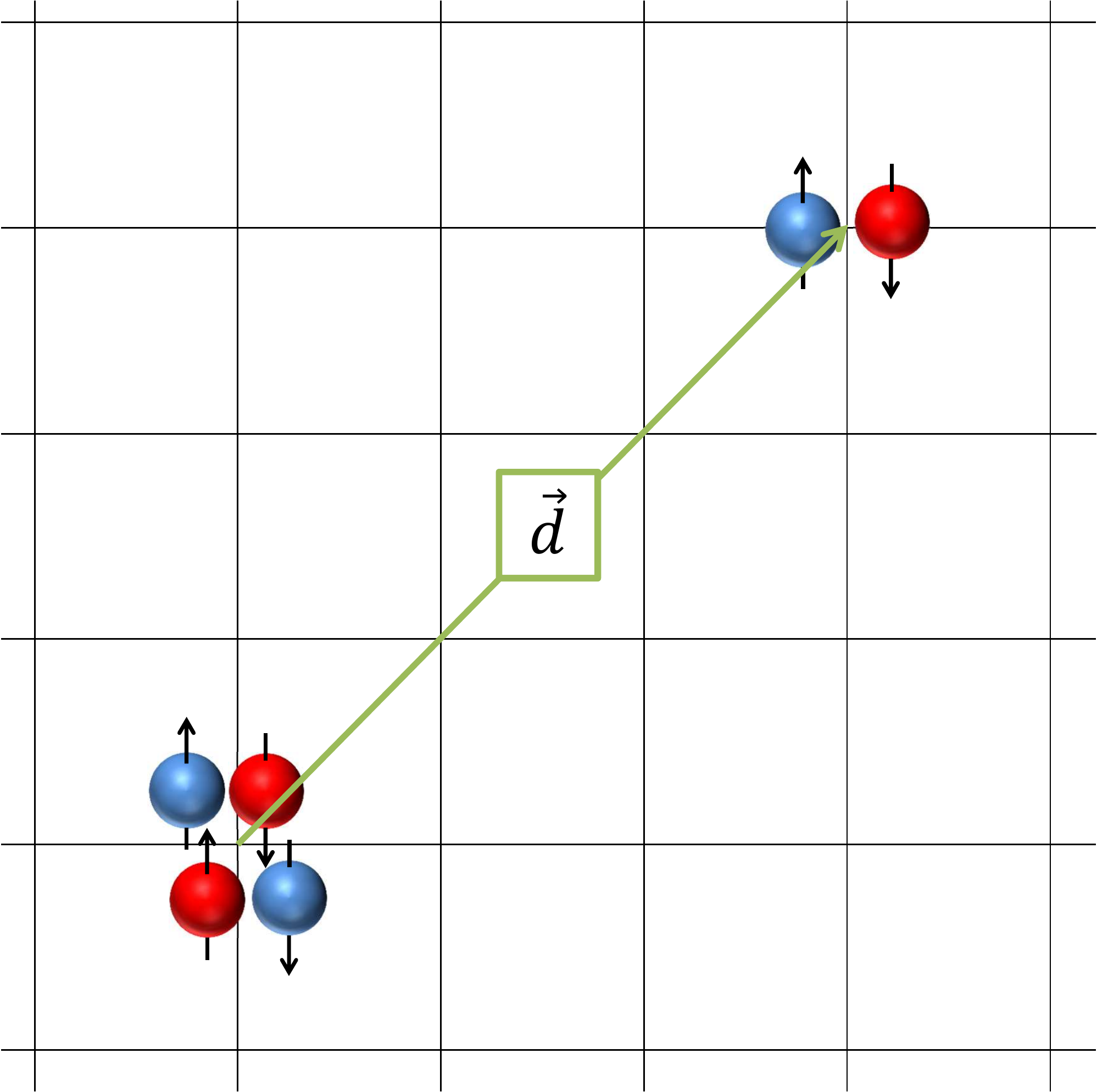}
\caption{A two-dimensional picture of the two-cluster initial state $\ket{\vec{d}\,\,}$ separated by displacement vector $\vec{d}$.}
\label{fig:initial-clusters}
\end{figure}

In this section we discuss the adiabatic projection method in detail and several new techniques to efficiently and accurately extract scattering information from the lattice calculations. To make our discussion general and complete, we consider two clusters consisting of $A_{1}$ and $A_{2}$ nucleons, respectively. The adiabatic projection method is based on defining the initial cluster states of clusters ${1}$ and ${2}$ on the lattice and evolving these initial cluster states in Euclidean time. The physical motivation is that we start with an approximate description of two cluster states, and then we use the microscopic interaction to evolve these states to the true low-lying cluster states in the presence of interactions. This has the advantage that the cluster states include all the deformation and polarization of the clusters due to the microscopic interaction, by design. We define the Slater-determinant initial cluster states as,
%------------ Equation -------------------------
\begin{align}
\ket{\vec{d}\,\,} = \sum_{\vec{n}} \ket{\vec{n}+\vec{d}\,\,}_{1} \otimes \ket{\vec{n}}_{2}
\,,
\label{eqn:InitialClusterStates-001}
\end{align}
where the states are parameterized by the two-cluster displacement vector $\vec{d}$. Fig.~\ref{fig:initial-clusters} is a schematic view of initial cluster state in two dimensions for two clusters separated by displacement vector $\vec{d}$.
%-----------  Figure ------------------
To improve the efficiency of the calculations, following Ref.~\cite{Elhatisari:2015iga}, we project the initial states onto spherical harmonics\footnote{This can be viewed as defining the cluster states in radial coordinates. To do so we bin the $\braket{n_{\hat{x}},n_{\hat{y}},n_{\hat{z}}}$ lattice points with the same distance $|\vec{d}|$ by weighting them with spherical harmonics ${\rm{Y}}_{\ell,\ell_{z}}(\hat{d})$.}
 ${\rm{Y}}_{\ell,\ell_{z}}$ with angular momentum quantum numbers $\ell$, $\ell_{z}$,
 %------------ Equation -------------------------
\begin{align}
 \ket{d}^{\ell,\ell_{z}} = 
 \sum_{\vec{d}^{\prime}}
 {\rm{Y}}_{\ell,\ell_{z}}(\hat{d}\,')
 \delta_{d,|\vec{d}\,'|}
 \ket{\vec{d}\,'} \,.
 \label{eqn:InitialClusterStates-005}
\end{align}
As we will discuss later, the systematic errors due to this projection are very small. This method was recently used in Ref.~\cite{Lu:2015riz} for scattering of two spin-$\frac{1}{2}$ particles with total spin $1$ on the lattice, and the scattering phase shifts and partial wave mixing angles were computed with high precision.

%--------------- Figure ---------------
\begin{figure}[!ht]
\includegraphics[width=0.4\textwidth]{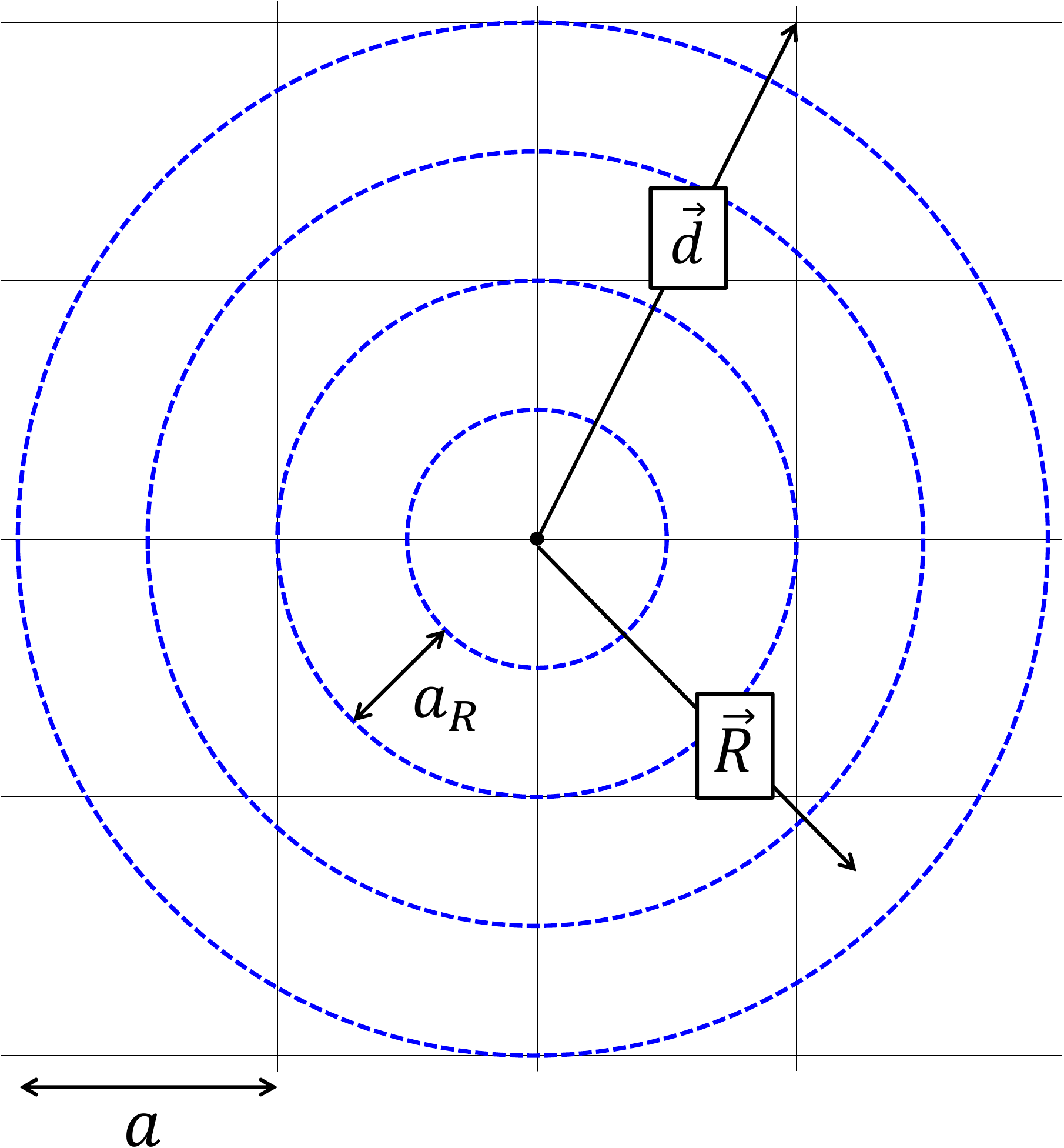}
\caption{A two-dimensional picture of the lattice and the radial bin size $a_{R}$ in the radial direction.}
\label{fig:radial-groups}
\end{figure}
To make the calculations more efficient, we find it convenient to define a radial bin size, $a_{R}$, in the radial direction and group together lattice points within the bin size $a_{R}$ in radial distance as shown in Fig.~\ref{fig:radial-groups}. As a result, we have grouped the lattice points into radial rings of width $a_{R}$. We now define $R$ as radial distance to the midpoint of the corresponding ring. Therefore, the initial cluster states are now
%------------ Equation -------------------------
\begin{align}
 \ket{R}^{\ell,\ell_{z}} = 
 \sum_{|d-R|<{a_{R}}/{2}}
 \ket{d}^{\ell,\ell_{z}}\,.
 \label{eqn:InitialClusterStates-009}
\end{align}

The methods introduced here provide significant improvements in the calculations with the adiabatic projection method and make the large scale calculations with Monte Carlo much more practical. The systematic errors due to redefining the radial coordinates have been analyzed by considering a two-particle system, and we have found that they are numerically very small. The two-particle scattering and the systematic errors are discussed in Section~\ref{appndx:two-particle-scattering}.  These methods have been recently used in the alpha-alpha scattering phase shift calculations~\cite{Elhatisari:2016owd}. In Section~\ref{sec:fermion-dimer-scattering}, we apply the methods described here to the three nucleon system to calculate the effective two-cluster interactions between neutron-deuteron ($n$-$d$) and proton-deuteron ($p$-$d$) in the spin quartet channel.

These initial cluster states satisfy the following completeness relation,
%------------ Equation -------------------------
\begin{align}
\mathbb{1} = \sum_{{R},{R}^{\prime}} 
\ket{R}^{\ell,\ell_{z}}
\
[N_{0}^{-1}]_{{R},{R}^{\prime}}^{\ell,\ell_{z}}
\
\prescript{\ell,\ell_{z}}{}{\bra{R^{\prime}}}
\,,
\label{eqn:CompletenesRel-001}
\end{align}
where $[N_{0}]_{R,R^{\prime}}^{\ell,\ell_{z}}$ is the norm matrix computed using the initial cluster states and given by
%------------ Equation -------------------------
\begin{align}
[N_{0}]_{R,R^{\prime}}^{\ell,\ell_{z}} = 
\prescript{\ell,\ell_{z}}{}{\braket{R|R^{\prime}}}^{\ell,\ell_{z}}
\,.
\label{eqn:NormMatrix-001}
\end{align}
The matrix elements are measures of the overlap between an initial cluster state with the clusters separated by $|\vec{R^{\prime}}|$ and a final cluster state with the clusters separated by $|\vec{R}|$. Now using the transfer matrix operator $\hat{M}$ we evolve the initial cluster states in Euclidean time to form the dressed cluster states,
%------------ Equation -------------------------
\begin{align}
\ket{R}^{\ell,\ell_{z}}_{{n_{t}}}
=
\hat{M}^{n_{t}}
\,
\ket{R}^{\ell,\ell_{z}}  
 \,,
\label{eqn:InitialClusterStates-013}
\end{align}
where $n_{t} = (L_{t}-1)/2$. Then we use these dressed cluster states to compute the norm matrix at a later Euclidean time $n_t$,
%------------ Equation -------------------------
\begin{align}
[N_{L_{t}}]_{R,R^{\prime}}^{\ell,\ell_{z}} = 
\sum_{R'',R'''}
[N_{0}^{-1/2}]_{R,R^{\prime\prime}}^{\ell,\ell_{z}}
\,
\prescript{\ell,\ell_{z}}{n_{t}}{\braket{R^{\prime\prime}|R^{\prime\prime\prime}}}_{n_{t}}^{\ell,\ell_{z}}
\,
[N_{0}^{-1/2}]_{R^{\prime\prime\prime},R^{\prime}}^{\ell,\ell_{z}}
\,,
\label{eqn:NormMatrix-005}
\end{align}
and the projected radial transfer matrix,
%------------ Equation -------------------------
\begin{align}
[M_{L_{t}}]_{R,R^{\prime}}^{\ell,\ell_{z}} = 
\sum_{R'',R'''}
[N_{0}^{-1/2}]_{R,R^{\prime\prime}}^{\ell,\ell_{z}}
\,
\prescript{\ell,\ell_{z}}{n_{t}}{\braket{R^{\prime\prime}
|\hat{M}|R^{\prime\prime\prime}}}_{n_{t}}^{\ell,\ell_{z}}
\,
[N_{0}^{-1/2}]_{R^{\prime\prime\prime},R^{\prime}}^{\ell,\ell_{z}}
\,.
\label{eqn:HamiltonianMatrix-001}
\end{align}
The radial transfer matrix given by Eq.~(\ref{eqn:HamiltonianMatrix-001}) already encodes the low-energy physics of the two-cluster system ($A_{1}$-body + $A_{2}$-body system). However, due to the fact that the dressed cluster states are not orthogonal, it is necessary to construct the norm matrix given by Eqs.~(\ref{eqn:NormMatrix-005}). Therefore,  we define the radial adiabatic transfer matrix as,
%------------ Equation -------------------------
\begin{align}
[M_{L_{t}}^{\text{a}}]_{R,R^{\prime}}^{\ell,\ell_{z}} =
\sum_{R'',R'''}
[N_{L_{t}}^{-1/2}]_{R^{\prime},R^{\prime\prime}}^{\ell,\ell_{z}}
\
[M_{L_{t}}]_{R^{\prime\prime},R^{\prime\prime\prime}}^{\ell,\ell_{z}}
\
[N_{L_{t}}^{-1/2}]_{R^{\prime\prime\prime},R^{\prime}}^{\ell,\ell_{z}} 
\,.
\label{eqn:AdiabaticHamiltonian-001}
\end{align}

Ref.~\cite{Rokash:2015hra} extensively discussed the adiabatic matrix in the asymptotic region where the two clusters are well-separated, and it was rigorously proven  that in the asymptotic region where the amount of overlap between the cluster wave packets is negligible, the adiabatic transfer matrix corresponds to an effective transfer matrix $M^{\text{eff}}$ that is simply a free (trivial) transfer matrix for two clusters along with any long-range interactions between the clusters. The important consequence of this is that the Euclidean time dependence of the adiabatic transfer matrix drops out in the asymptotic region, and Eq.~(\ref{eqn:AdiabaticHamiltonian-001}) reads
%------------ Equation -------------------------
\begin{align}
[M_{L_{t}}^{\text{a}}]_{R,R^{\prime}}^{\ell,\ell_{z}} =
[M^{\text{eff}}]_{R,R^{\prime}}^{\ell,\ell_{z}}
\, ,
\label{eqn:AdiabaticHamiltonian-005}
\end{align}
where $R$ or $R'$ corresponds to the indices with  distances larger than the short-range interaction scale.

%---------------- Figure ---------------
\begin{figure}[!ht]
\centering
\begin{minipage}[b]{0.32\textwidth}
\includegraphics[clip,trim=1.5cm 1cm 0cm 0cm,width=\textwidth]{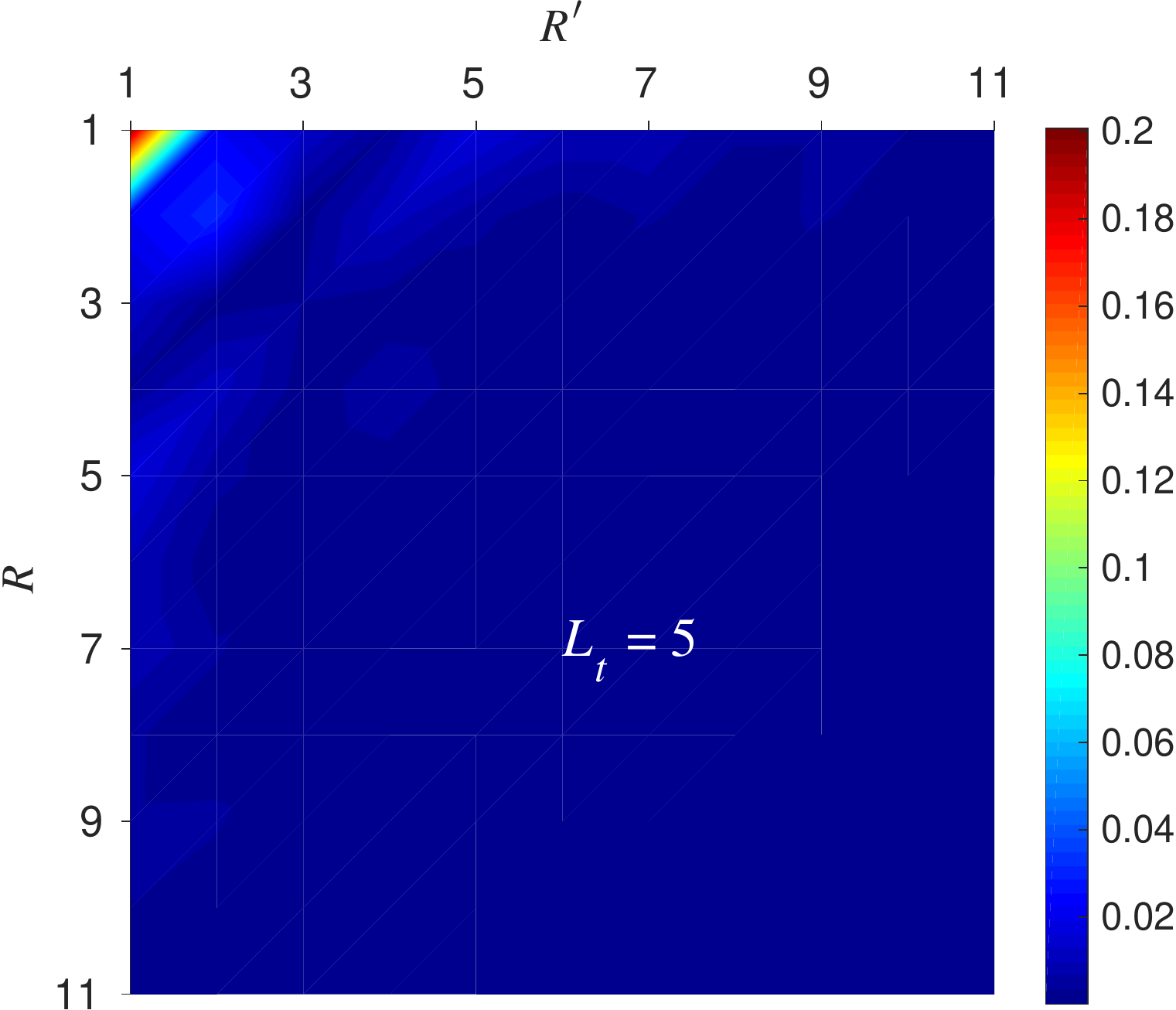}
\end{minipage}
\begin{minipage}[b]{0.32\textwidth}
\includegraphics[clip,trim=1.5cm 1cm 0cm 0cm,width=\textwidth]{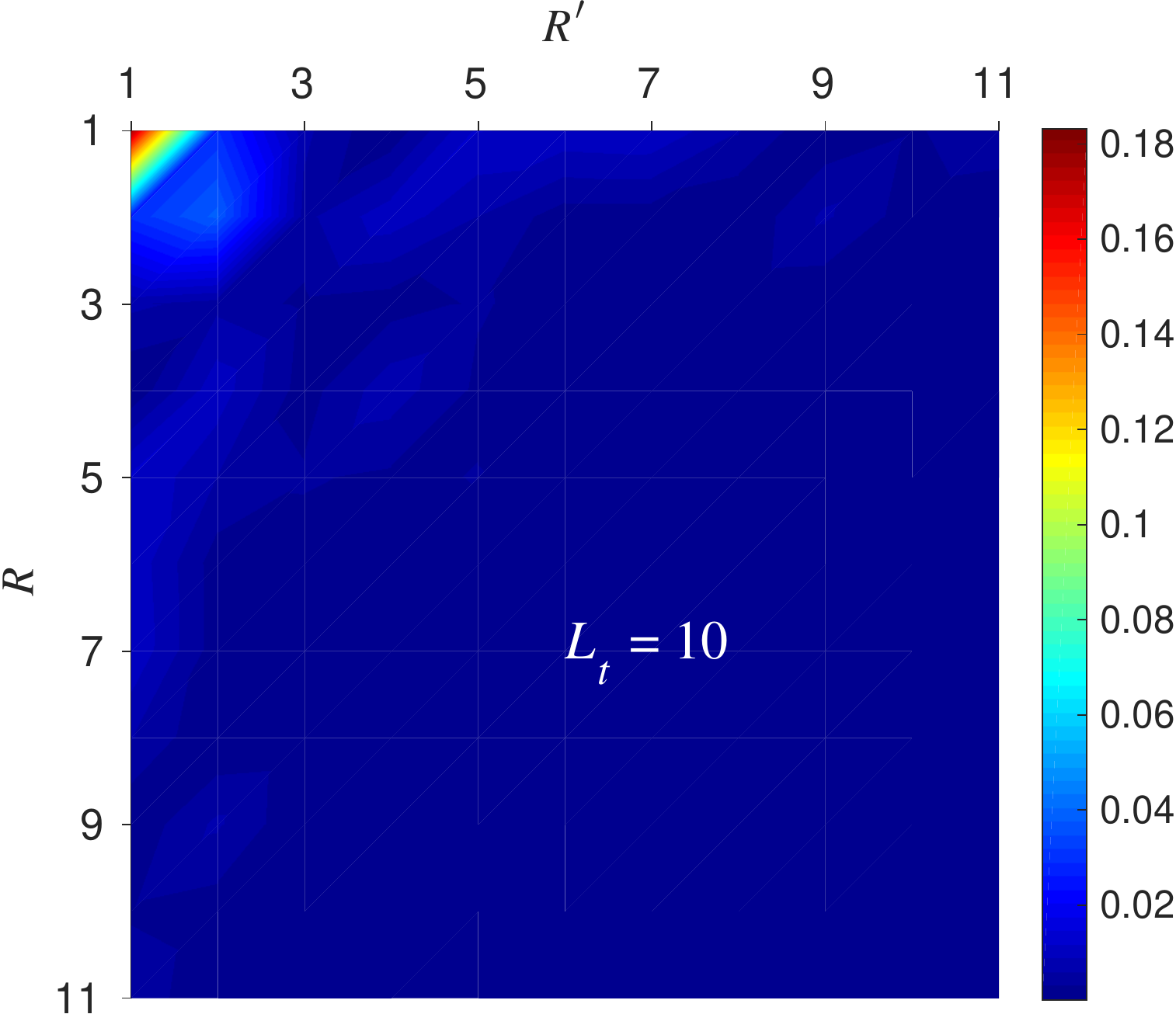}
\end{minipage}
\begin{minipage}[b]{0.32\textwidth}
\includegraphics[clip,trim=1.5cm 1cm 0cm 0cm,width=\textwidth]{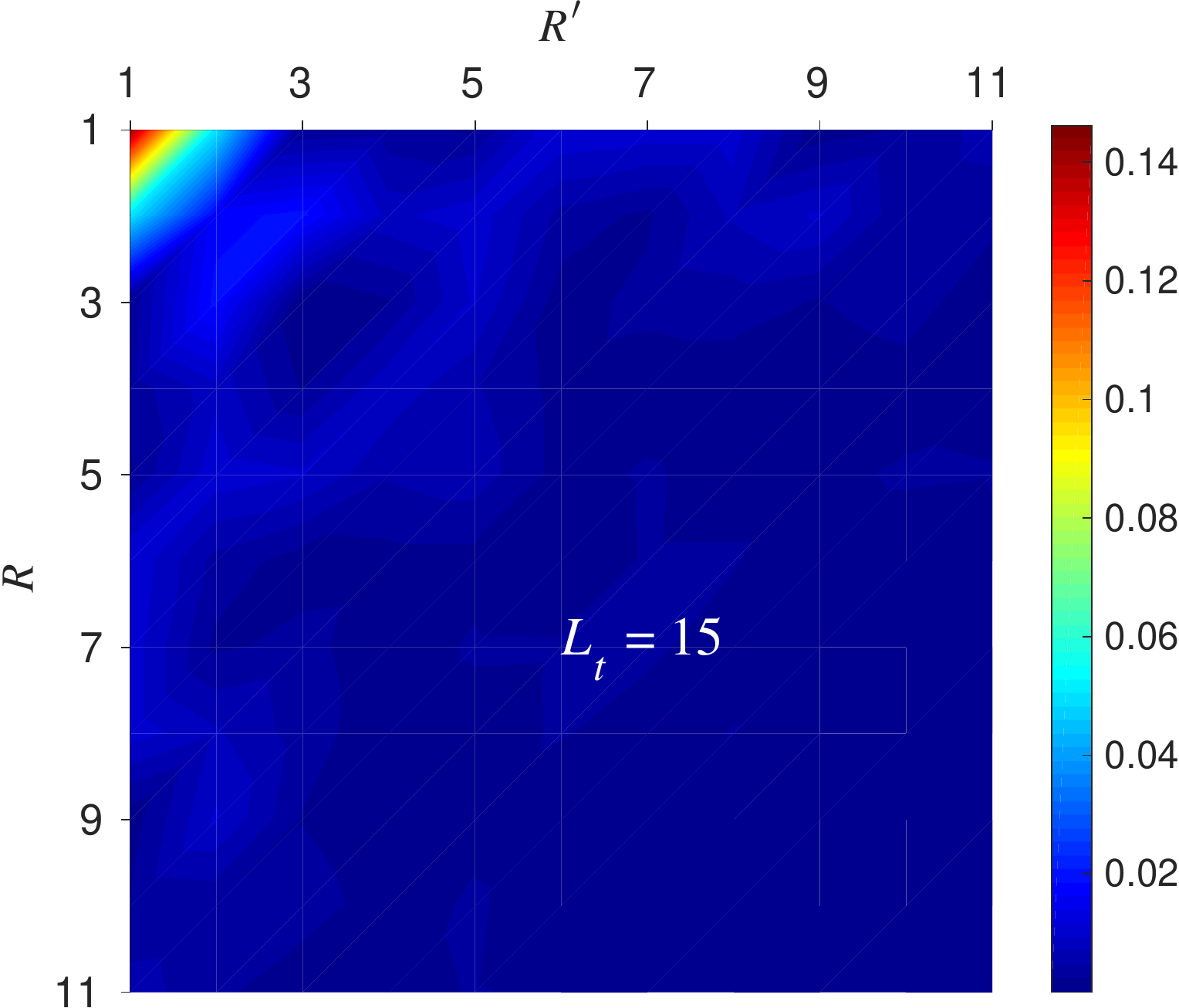}
\end{minipage}
\caption{The absolute difference between $[M_{L_{t}}^{\text{a}}]_{R,R^{\prime}}^{\ell,\ell_{z}}$ and $[M^{\text{eff}}]_{R,R^{\prime}}^{\ell,\ell_{z}}$ for fermion-dimer system with the angular quantum numbers $\ell = 0$ and $\ell_{z}=0$ and for the Euclidean time steps $L_{t} =$ 5, 10, and 15, respectively. The absolute difference matrices are in lattice units. For $L_{t}=15$ the magnitude of the absolute difference is equal to $18\%$ of the corresponding value of the trivial radial adiabatic transfer matrix for the lattice site where the absolute difference is maximum. The physical distances for the indices $R$, $R'=$ 1, 3, 5, 7, 9, and 11 are 2.22, 3.20, 4.68, 6.16, 7.14, and  8.62~fm, respectively.  }
\label{fig:Hamiltonian-difference}
\end{figure}

In Fig.~\ref{fig:Hamiltonian-difference} we show the absolute difference between the interacting two-cluster adiabatic transfer matrix and the trivial two-cluster adiabatic transfer matrix for fermion-dimer system. The fermion-dimer system will be discussed in Section~\ref{sec:fermion-dimer-scattering}. Here we point out that at distances less than $6$~fm we see the effects of the shallow deuteron bound state. The bound state deuteron wave function is characterized by a length scale $~1/\kappa$ where  the binding momentum $\kappa\approx 45.7$ MeV.  At  distances larger than $1/\kappa$ the absolute difference approaches  zero as expected by Eq.~(\ref{eqn:AdiabaticHamiltonian-005}). This is an important result. It allows us to construct a two-cluster Hamiltonian in a periodic box that is only slightly larger than the range of the interaction. Then using  Eq.~(\ref{eqn:AdiabaticHamiltonian-005}) we can embed the lattice Hamiltonian in a much larger box that describes trivial evolution of the two clusters which is not as computationally demanding. It is important to be able to describe the interaction in the asymptotic region especially in the presence of the long-range Coulomb forces to meaningfully define a cross section.

\begin{figure}[!ht]
\centering{
\includegraphics[width=0.6\textwidth]{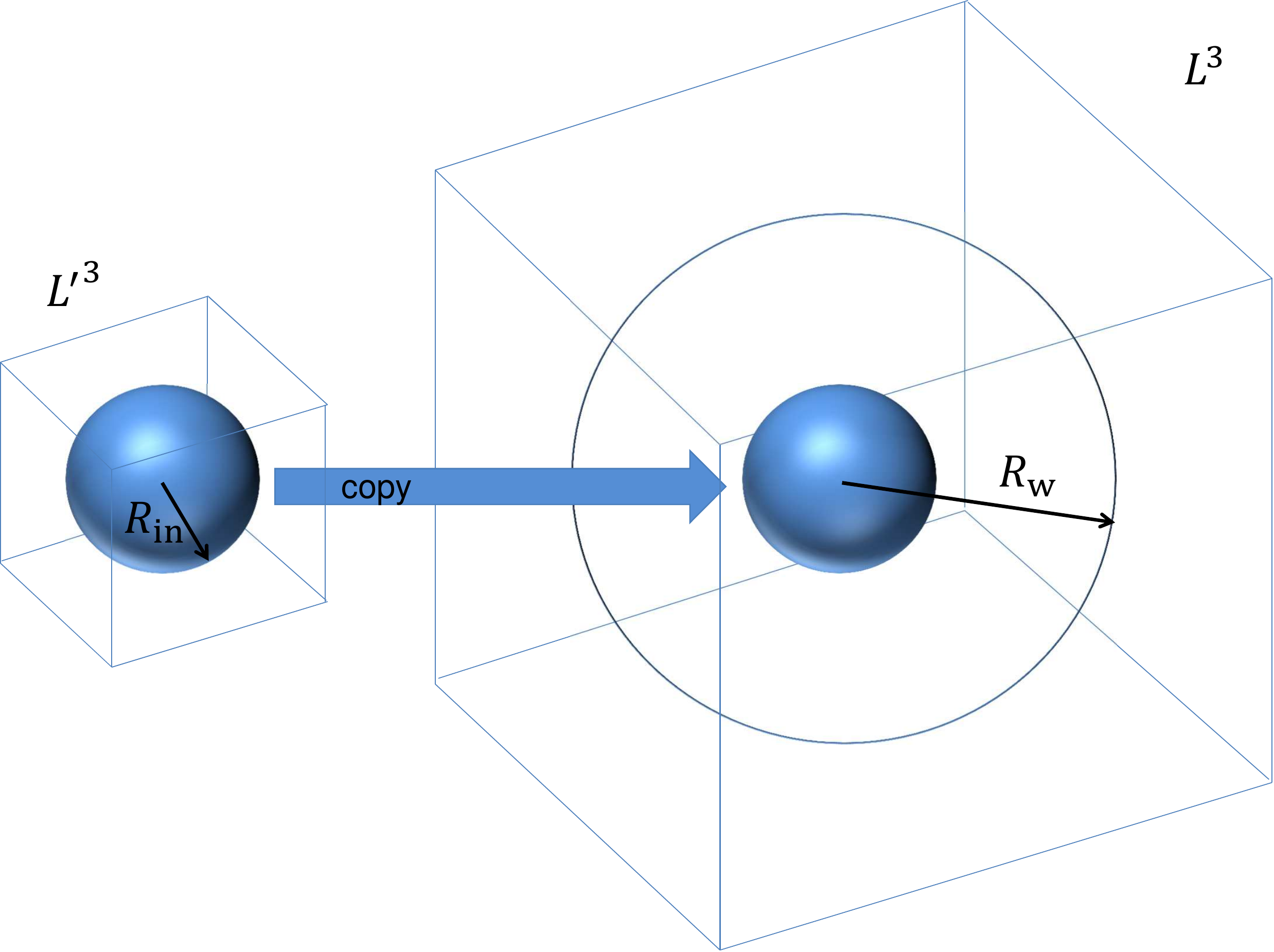}}
\caption{A sketch of the lattices for the cluster-cluster calculations in the overlapping and in the non-interacting regions. $R_{\text{in}}$ is the largest radial distance where $[M_{L_{t}}^{\text{a}}]_{R,R^{\prime}}^{\ell,\ell_{z}}$ and $[M^{\text{eff}}]_{R,R^{\prime}}^{\ell,\ell_{z}}$ are connected without introducing any systematic errors due to the periodic boundary of $[M_{L_{t}}^{\text{a}}]_{R,R^{\prime}}^{\ell,\ell_{z}}$ constructed in the cubic box with volume ${L'} ^{3}$. $R_{\text{w}}$ indicates the radius of the spherical wall discussed in Section~\ref{sec:Scattering-phase-shifts}.}
\label{fig:two_cluster_simulation}
\end{figure}

The calculation of the two-cluster Hamiltonian is divided into two parts to take advantage of the result from Fig.~\ref{fig:Hamiltonian-difference}. First, we 
compute the radial adiabatic transfer matrix given by Eq.~(\ref{eqn:AdiabaticHamiltonian-001}) for the $A_{1}$-body + $A_{2}$-body system. Due to the computational demand associated with $A_{1}$+$A_{2}$ active nucleons, this is done on a  periodic lattice with volume ${L\,'}^{3}\times L_{t}$ whose size is larger than the range of the interaction but not too large. Subsequently, we also compute the transfer matrices for individual clusters ${1}$ and ${2}$ on periodic cubic lattices where  all interactions between $A_{1}$ nucleons in the cluster $1$ and between $A_{2}$ nucleons in the cluster $2$ are turned on.  However, the inter-cluster interaction is turned off. 
This second calculation is done in a large  periodic lattice with volume $L^{3}\times L_{t}$ where $L\gg L'$. This calculation that describes  trivial two-cluster radial adiabatic transfer matrix of non-interacting clusters ${1}$ and ${2}$ along with any long-range Coulomb interactions becomes exact in the asymptotic region,
Eq.~(\ref{eqn:AdiabaticHamiltonian-005}).  As a corollary, we connect the interacting radial adiabatic transfer matrix with the trivial radial adiabatic transfer matrix in the asymptotic region to extend the interacting radial transfer matrix to the volume $L^{3}\times L_{t}$. This procedure is illustrated in Fig.~\ref{fig:two_cluster_simulation}. The periodic boundary of the ${L\,'}^{3}$ volume must be treated with care to avoid systematic errors. We only use the matrix elements of $[M_{L_{t}}^{\text{a}}]_{R,R^{\prime}}^{\ell,\ell_{z}}$ with $R, R\,' < L\,'/2$.  As shown in Fig.~\ref{fig:Hamiltonian-difference}, in the asymptotic region the absolute difference $|\, [M_{L_{t}}^{\text{a}}]_{R,R^{\prime}}^{\ell,\ell_{z}} -
[M^{\text{eff}}]_{R,R^{\prime}}^{\ell,\ell_{z}}\,|$ is very small, and the systematic errors due to extension of the interacting radial transfer matrix are under control. 

%--------------- Figure ---------------------

\begin{table}[!ht]
\caption{The size of the two-cluster adiabatic transfer matrices. The matrix $\[M^{\text{a}}_{L_{t}}\]_{\vec{d},\vec{d}'}$ is constructed from the initial cluster states on a cubic lattice. The matrix $\[M_{L_{t}}^{\text{a}}\]_{d,d^{\prime}}^{\ell,\ell_{z}}$ is obtained by projecting the initial cluster states onto spherical harmonics with the angular momentum quantum numbers $\ell = 0$, $\ell_{z}=0$. The matrix $\[M_{L_{t}}^{\text{a}}\]_{R,R^{\prime}}^{\ell,\ell_{z}}$ is constructed by grouping the lattice points into radial rings of width $a_{R}$ as well as the projection onto spherical harmonics.}
\label{table:comp-scaling}
\centering
\begin{tabular}{@{}lcccc@{}}
\hline\hline
\multirow{2}{*}{$~L~$}
&\multirow{2}{*}{$\[M^{\text{a}}_{L_{t}}\]_{\vec{d},\vec{d}'}$} 
&\multirow{2}{*}{$\[M_{L_{t}}^{\text{a}}\]_{d,d^{\prime}}^{0,0}$}
& \multicolumn{2}{c}{$\[M_{L_{t}}^{\text{a}}\]_{R,R^{\prime}}^{0,0}$}
\\
\cmidrule[0.2pt](lr{0.5em}){4-5}
& \multicolumn{1}{c}{\qquad \qquad}
& \multicolumn{1}{c}{\qquad \qquad}
& \multicolumn{1}{c}{$a_{R} = 0.125$~\text{l.u.}}  \quad
& \multicolumn{1}{c}{$a_{R} = 0.250$~\text{l.u.}}  \quad
\\\hline
~10~
& ~$10^3\times 10^3$~
& ~$22\times 22$~
& ~$21\times 21$~
& ~$14\times 14$~
\\\hline
~20~
& ~$20^3\times 20^3$~
& ~$85\times 85$~
& ~$58\times 58$~
& ~$34\times 34$~
\\\hline
~30~
& ~$30^3\times 30^3$~
& ~$189\times 189$~
& ~$97\times 97$~
& ~$54\times 54$~
\\\hline
~40~
& ~$40^3\times 40^3$~
& ~$335\times 335$~
& ~$137\times 137$~
& ~$74 \times 74$~
\\\hline
~50~
& ~$50^3\times 50^3$~
& ~$522\times 522$~
& ~$177\times 177$~
& ~$94 \times 94$~
\\\hline
~60~
& ~$60^3\times 60^3$~
& ~$752\times 752$~
& ~$217\times 217$~
& ~$114 \times 114$~
\\\hline\hline   
\end{tabular}
\end{table}
Before ending this section, in Table~\ref{table:comp-scaling} we show how the dimensions of the adiabatic transfer matrix scales after projection onto spherical harmonics and grouping the lattice points into radial rings of width $a_{R}$. The numbers indicated here are valid for any two-cluster system, and they are independent of $A_{1}$ and $A_{2}$. Therefore, Table~\ref{table:comp-scaling} clearly shows that the new tools introduced in this section make the large scale calculations with Monte Carlo much more practical and efficient. Recently, the alpha-alpha scattering phase shifts have successfully been computed from the radial adiabatic transfer matrix using this technique with $a_{R} = 0.25$~l.u.~\cite{Elhatisari:2016owd}.

%=================================   Spherical Wall Method =====================
\section{Scattering phase shifts from a finite volume}
\label{sec:Scattering-phase-shifts}
%==========================================================================

In the previous section, we constructed the adiabatic transfer matrix in radial coordinates, and the improvement in the computational scaling is summarized in Table~\ref{table:comp-scaling}. To extract the scattering information from the adiabatic transfer matrix defined in the radial coordinates, we use the so-called spherical wall method~\cite{Carlson:1984zz,Borasoy:2007vy} which gives us access to the scattering phase shift of the two-cluster system.

\begin{figure}[!ht]
\centering
\begin{minipage}[b]{0.43\textwidth}
\includegraphics[width=0.89\textwidth]{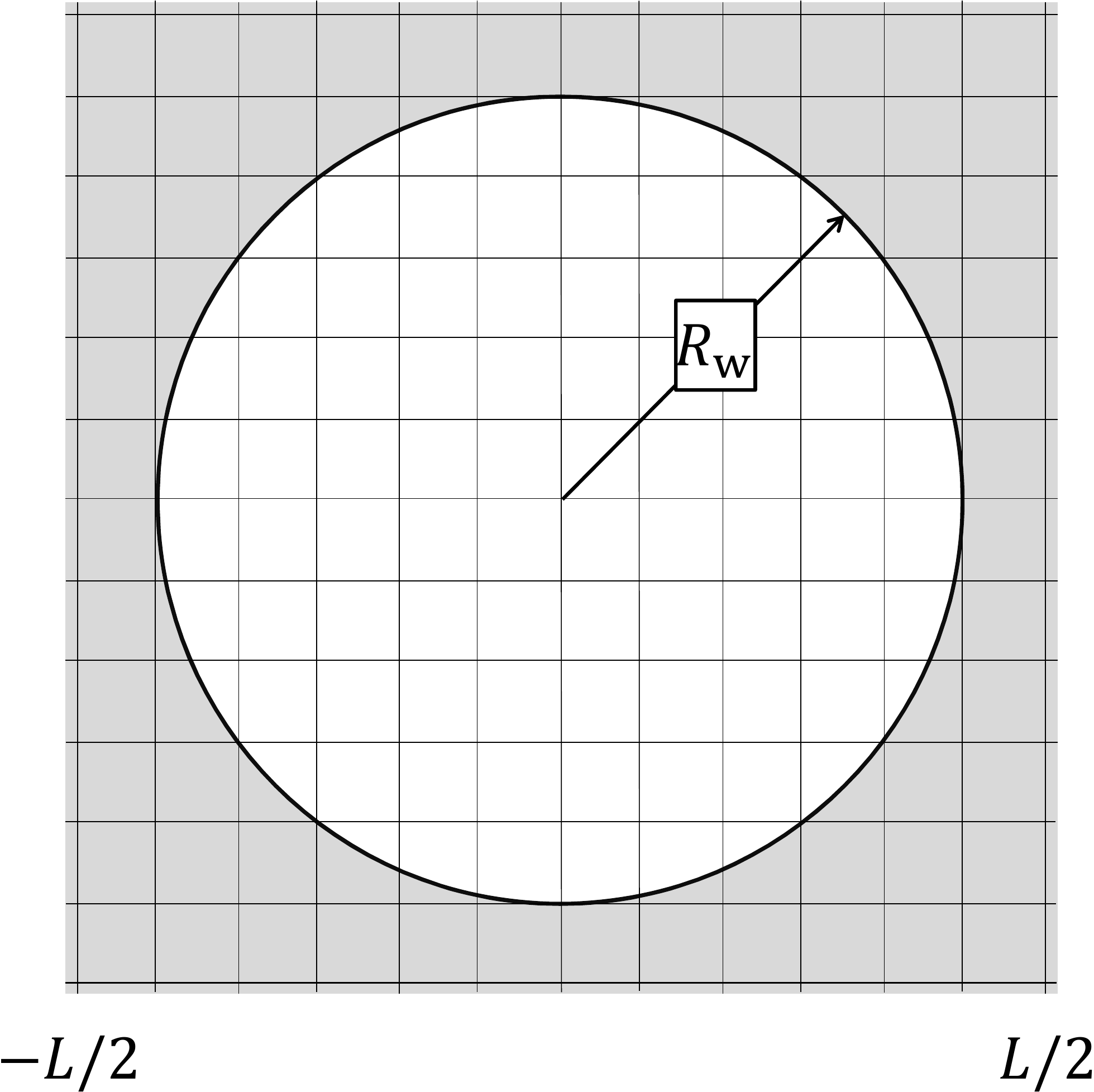}
\vspace{-0.15cm}
\end{minipage}
\hspace{0.9cm}
\begin{minipage}[b]{0.44\textwidth}
\includegraphics[width=\textwidth]{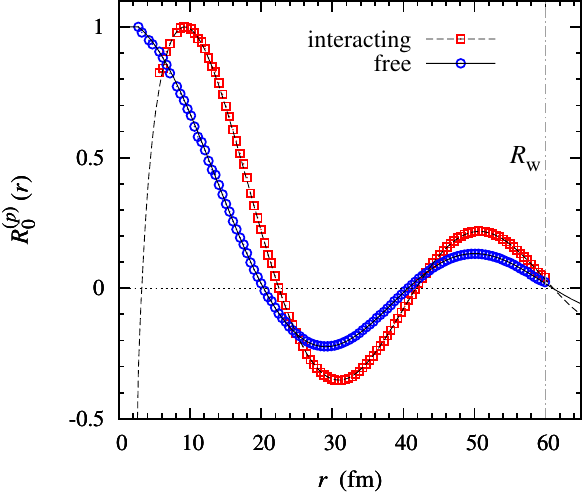}
\end{minipage}
\caption{(Left panel) A schematic view of the spherical wall with radius $R_{\text{w}}$ imposed on two-dimensional lattice. (Right panel) The plots of the $S$-wave scattering wave functions from the interacting and trivial two-cluster radial adiabatic transfer matrices.}
\label{fig:hard-wall}
\end{figure}

The spherical wall method imposes a hard boundary wall condition on the relative separation between two clusters in finite box and removes the periodic boundary effects inherited from the cubic lattice. This eliminates artifacts due to the periodic boundary and allows one to accurately compute scattering phase shifts at higher
orbital angular momenta and to study systems with the spin-orbital couplings~\cite{Borasoy:2007vy}. Further, the spherical hard wall is a general method to extract the phase shifts in the presence of a long-range interaction and at energies above the inelastic threshold. These are important considerations for lattice calculations. For example, there is no non-perturbative treatment of the Coulomb force in  L\"uscher's method~\cite{Luscher:1986pf,Luscher:1990ux} but this is easily done in using the spherical wall formulation. Fig.~\ref{fig:hard-wall}(a) is a schematic view of a spherical wall at radius $r = R_{\text{w}}$ in two dimensions. Solving the Schr\"odinger equation for the radial adiabatic transfer matrix of the two-cluster system with the imposed spherical hard wall, we obtain the spherical scattering waves which vanish at $R_{\text{w}} +\varepsilon_{R}$, where $\varepsilon_{R}$ is the correction on the precise radius of the spherical wall due to the discrete lattice, and it is less than a half of the lattice spacing, $|\varepsilon_{R}| < a/2$. In Fig.~\ref{fig:hard-wall}(b) we show an example of the spherical scattering lattice wave functions for trivial and interacting fermion-dimer systems.
%-----------  Figure ------------------

Following Refs.~\cite{Borasoy:2007vy,Rokash:2015hra}, in the asymptotic region we match the lattice wave function with the spherical scattering wave in continuum and extract the scattering phase shifts with high accuracy. The total wave function in continuum is decomposed into the radial
part $R_{\ell}^{(p)}$ and spherical harmonics $Y_{\ell,\ell_{z}}(\hat{r})$,
\begin{align}
\Psi(\vec{r})
=
R_{\ell}^{(p)}(r) \, \text{Y}_{\ell,\ell_{z}}(\hat{r})\,,
\label{eqn:asymptotic-wave-functions-001}
\end{align}
where $p$ is the relative momentum of the clusters. The asymptotic form of the radial wave function  $R_{\ell}^{(p)}$ is
\begin{align}
R_{\ell}^{(p)}(r) =
N_{\ell}(p) \times
\begin{dcases}
\cot\delta_{\ell}(p) \, j_{\ell}(p\,r) - n_{\ell}(p\,r) &  \quad \text{for any finite-range potential,} \\
\cot\delta_{\ell}(p) \, F_{\ell}(p\,r) + G_{\ell}(p\,r)  & \quad  \text{for charged clusters,}\,
\end{dcases}
\label{eqn:Rwall-009}
\end{align}
where $N_{\ell}(p)$ is an overall normalization coefficient, $j_{\ell}~(n_{\ell})$ is spherical Bessel function of the first (second) kind, and $ F_{\ell}~(G_{\ell})$ is the regular (irregular) Coulomb wave function~\cite{Abramowitz,Koenig:2012bv}. Since our lattice wave functions emerge from the solutions of the Schr\"odinger equation for the radial transfer matrix projected onto spherical harmonics with the angular momentum quantum numbers $\ell$ and $\ell_{z}$, we can extract the scattering phase shifts by fitting Eq.~(\ref{eqn:Rwall-009}) to the lattice wave functions. However, as we explain in the following, we find that working with the spectrum of the radial adiabatic transfer matrix and the trivial radial transfer matrix is more efficient and accurate than the wave function fitting procedure.

%-----------  Figure ------------------
\begin{figure}[!ht]
\centering
\begin{minipage}[b]{0.45\textwidth}
\includegraphics[width=\textwidth]{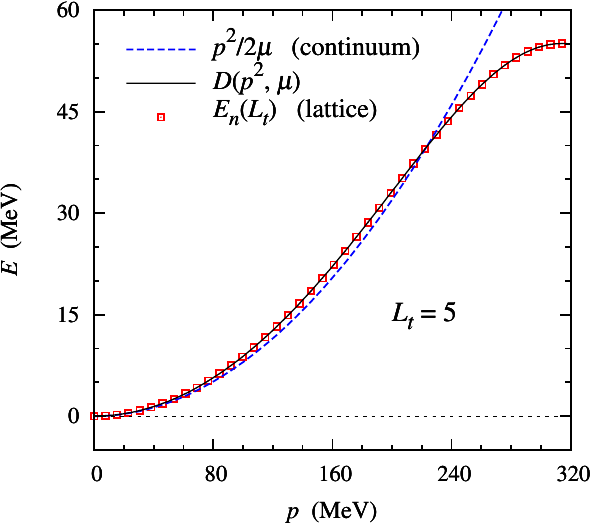}
\end{minipage}
\hfill
\begin{minipage}[b]{0.45\textwidth}
\includegraphics[width=\textwidth]{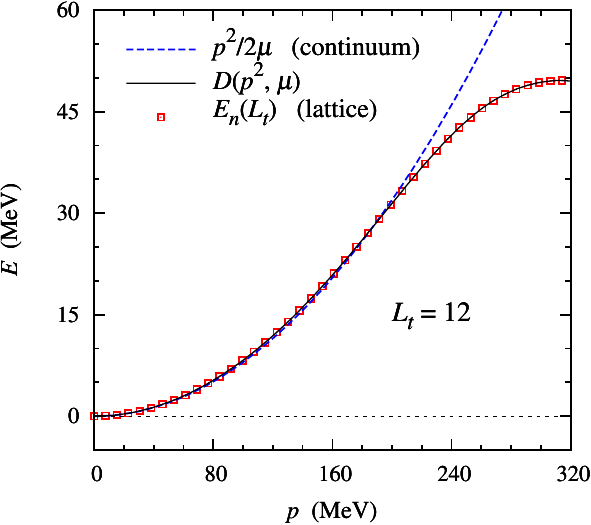}
\end{minipage}
\caption{The fermion-dimer dispersion relations for the $\mathcal{O}(a^{4})$-improved lattice action and $\pi a^{-1}\simeq$~314 MeV. Here the fermion-dimer reduced mass is $\mu =$~626 MeV.}
\label{fig:dispersio-relation}.
\end{figure}
The spectrum of the trivial radial adiabatic transfer matrix provides crucial information to calculate the correction $\varepsilon_{R}$ to the radius of the spherical hard wall. For this case the system has zero phase shift and the wave function vanishes at $R^{\,\prime}_{\text{w}} = R_{\text{w}} + \epsilon_{R}$. Using the corrected radius of the spherical hard wall and the spectrum of the interacting system, we solve Eq.~(\ref{eqn:Rwall-009}) for the scattering phase shifts,
\begin{align}
\delta_{\ell}(p) =
\begin{dcases}
 \tan^{-1}\[\frac{j_{\ell}(p\,R^{\,\prime}_{\text{w}})}{n_{\ell}(p\,R^{\,\prime}_{\text{w}})}\] &  \quad \text{for any finite-range potential,} \\
-\tan^{-1}\[\frac{ F_{\ell}(p\,R^{\,\prime}_{\text{w}})}{G_{\ell}(p\,R^{\,\prime}_{\text{w}})}\]  & \quad  \text{for charged clusters.}\,
\end{dcases}
\label{eqn:phase-shifts-001}
\end{align}
The relative momentum $p$ is computed from the spectrum of the radial adiabatic transfer matrix and the dispersion relation of the two-cluster system given by
\begin{align}
E = D(p^{2},\mu) = c_{0}\,\frac{p^{2}}{2\mu} + c_{1}\,p^{4} + c_{2}\,p^{6} + \ldots  \,.
\label{eqn:dispersion-relation}
\end{align}
where $\mu$ is the reduced mass of the two-cluster system, and the coefficients $c_{i}$ are determined by fitting Eq.~(\ref{eqn:dispersion-relation}) to the lattice dispersion relation. In Fig.~\ref{fig:dispersio-relation} we show the lattice and continuum dispersion relations for a two-cluster system consisting of a fermion and dimer at two different Euclidean times.

The energy Eq.~(\ref{eqn:dispersion-relation}) is calculated from the lattice using that the eigenvalues $\lambda_{n}(L_{t})$ of any transfer matrix is related to the energy estimate at $L_{t}$ time steps. This relation is given by
\begin{align}
 E_{n}(L_{t}) = -\frac{\log\[{\lambda_{n}(L_{t})}\]}{\alpha_{t}}\, .
\label{eqn:spectrum-001}
 \end{align}
The energy $E_{n}$ is computed by extrapolation to the limit $L_{t}\to\infty$ using the ansatz,
\begin{align}
E_{n}(L_{t}) = 
E_{n} + c_{n}(E) \, e^{-L_{t} \alpha_{t} \, \Delta E_{n}}\,.
\label{eqn:spectrum-005}
\end{align}
This ansatz is valid when the splittings from the excited states are large. We discuss the general case in Appendix~\ref{sec:extrapolation}.  We can also do a similar infinite time extrapolation of the two-cluster scattering phase shift calculated from $E(L_t)$  in Eq.~(\ref{eqn:phase-shifts-001}) using a similar anzatz:
\begin{align}
\delta_{\ell}[p(L_{t})] = \delta_{\ell}(p)  + c_{\ell}(p) \, e^{-L_{t} \alpha_{t}\, \Delta E_{\ell}}\,.
\label{eqn:phase-shifts-005}
\end{align}
For a detailed discussion of Eq.~(\ref{eqn:phase-shifts-005}) see Appendix~\ref{sec:extrapolation}. The extrapolations in Eqs.~(\ref{eqn:spectrum-005}) and (\ref{eqn:phase-shifts-005}) eliminate the contributions from excited states. We have found that the final results for the scattering phase shifts are independent of which extrapolation is adopted in the analysis. However, since the phase shift extrapolation is more efficient, the final results for the scattering phase shifts in this work are obtained by extrapolating the phase shifts.

%========================== Two-particle Scattering =============================
\section{Two-particle scattering on the lattice}
\label{appndx:two-particle-scattering}
%===========================================================================

In this section, we analyze and discuss the systematic errors due to the methods introduced Section~\ref{sec:adiabatic-projection} for the calculations of the adiabatic transfer matrix. To check the systematic errors arising from the projection onto spherical harmonics and the radial bin size $a_{R}$, we first consider a simple system of two distinguishable  spin-$\frac{1}{2}$ particles with equal masses on a periodic cubic lattice. Here we use the adiabatic projection formalism with $L_{t} = 1$ to also check errors associated with extending the volume from ${L\,'}^{3}$ to ${L}^{3}$. We perform the calculations with lattice volumes ${L'}^{3} = (16~\text{l.u.})^{3}$ and ${L}^{3} = (62~\text{l.u.})^{3}$.

We consider an attractive Gaussian potential between the particles,
\begin{align}
V(r)  = c_{g} \exp\[-\frac{r^{2}}{2 R_{g}^{2}}\]
\label{eqn:Gaussian-potential}
\end{align}
where the interaction strength $c_{g}$ and the range of the potential $R_{g}$ are set to $-8.0$ MeV and $0.02$~MeV$^{-1}$, respectively. We use the same free lattice Hamiltonian as given in Eq.~(\ref{eqn:free-Hamiltonian-002}). Therefore, the microscopic transfer matrix for the two-particle system is 
\begin{align}
\hat{M}
=
:
\exp\[- \alpha_{t}\,\hat{H}_{0}
-\alpha_{t}\, c_{g}\,
\sum_{\vec{n}}
\exp\[-\frac{|\vec{n}|^{2}}{2 R_{g}^{2}}\]
\rho_{\spinup}(\vec{n})
\rho_{\spindown}(\vec{n})
\]
:\,.
\label{eqn:Transfer-matrix-101}
\end{align}
%-----------  Figure ------------------
\begin{figure}[!ht]
\includegraphics[width=0.96\textwidth]{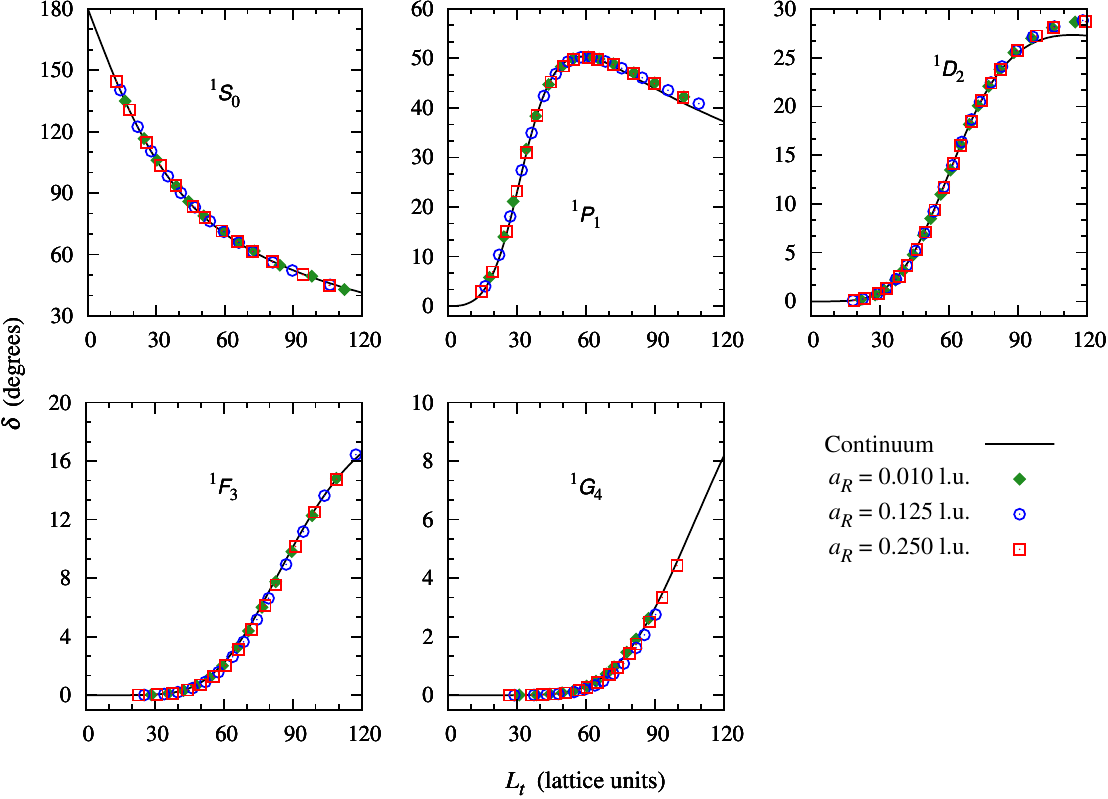}
\caption{Plots of the two-particle scattering phase shifts for the partial waves with $\ell \leq 4$ versus relative momentum. The two-particle system has total spin 0 and the interaction between particles is an attractive Gaussian potential. The scattering phase shifts are computed from the radial adiabatic transfer matrices for $L_{t} = 1$ and with $a_{R}$= 0.010, 0.125, and 0.250 in lattice units~(l.u.), and projected onto spherical harmonics with $\ell \leq 4$ and $\ell_{z} =0$.}
\label{fig:two-body-phase-shifts}
\end{figure}

In Fig.~\ref{fig:two-body-phase-shifts} we show the two-particle scattering phase shifts for angular momenta  $\ell \leq 4$ and various radial bin size $a_{R}$ = 0.010~l.u., 0.125~l.u., and 0.250~l.u.. As can be seen from the scattering phase shifts, projecting onto spherical harmonics as given in Eq.~(\ref{eqn:InitialClusterStates-005}) and grouping the lattice points within the width $a_{R}$ introduce very small errors, at most a couple percent of relative error in the phase shift.
% We also note here that for $a_{R} \leq 0.02$~fm Eq.(\ref{eqn:InitialClusterStates-009}) reduces to Eq.(\ref{eqn:InitialClusterStates-005}).

%========================  Nucleon-Deuteron Scattering ================
\section{neutron-deuteron and proton-deuteron scattering}
\label{sec:fermion-dimer-scattering}
%===============================================================

In this section we consider fermion-dimer system consisting of two-component fermions. We denote the two components as $\spinup$ (spin up) and $\spindown$ (spin down). The fermion-dimer system is the simplest nontrivial  two-cluster problem one can study. We use it to test the methods developed and discussed in Sections~\ref{sec:adiabatic-projection} and \ref{sec:Scattering-phase-shifts}. There are several reasons that make the fermion-dimer system attractive for a benchmark lattice calculation.  At low energy this system can be described by an EFT with zero-range two-fermion interaction, and yet still have a very shallow dimer characterized by a large extended deuteron wave function. Thus on one hand, the lattice short-ranged interaction  is simple  to implement and on the other hand there is a finite length scale associated with the dimer over which we can test the lattice method of matching the fully interacting  $A_1+A_2$ adiabatic calculation to the $A_1$ and $A_2$ trivial adiabatic calculation as described in Section~\ref{sec:adiabatic-projection}, see Fig.~\ref{fig:two_cluster_simulation}. Moreover,  the continuum EFT calculation in this theory with and without Coulomb interaction is easily available~\cite{Bedaque:1997qi,Bedaque:1998mb,Gabbiani:1999yv,Bedaque:2002yg,Rupak:2001ci,Konig:2011yq}. This allows a direct comparison of the lattice data with the continuum calculations that both use the same low-energy theory with the same number of physics input --- in this case the deuteron binding energy.

The fermion-dimer problem can be related to nucleon-deuteron scattering in the spin-quartet channel. In the quartet channel when one considers the situation with all the spins aligned along the quantization axis, the isospin label that distinguishes a proton from  a neutron can be identified with the spin up $\spinup$  and and down 
$\spindown$  label of the fermion-dimer system, as the nucleon spin label becomes redundant. Thus quartet channel neutron-deuteron scattering is identified   with $\spindown$-dimer scattering, and proton-deuteron scattering with $\spinup$-dimer scattering. The dimer is a bound state of $\spinup$ and $\spindown$ fermion.

For simplicity, we set the mass of two-component fermions to be equal, $m_{\spinup} = m_{\spindown} = m$, and $m =$ 939~MeV. With the zero-range interactions between two-component fermions, the potential is written in the following form, 
\begin{align}
\hat{V} 
= C_{0}\,
\sum_{\vec{n}}
\rho_{\spinup}(\vec{n})
\rho_{\spindown}(\vec{n})
+
\alpha_{\text{EM}}\,
Z_{\spinup}^{2}\,
\sum_{\vec{n}}
\frac{\rho_{\spinup}(\vec{n})\rho_{\spinup}(\vec{n})}
{|\vec{n}|}\,,
\label{eqn:Interaction-005}
\end{align}
where $C_{0}$ is the lattice-regularized short-range interaction coefficient to be determined, $\alpha_{\text{EM}}=e^{2}/4\pi = 1/137.036$ is the fine-structure constant, $Z_{\spinup}$ is the charge of $\spinup$ component fermion. The Coulomb part of the potential does not require any special treatment at $|\vec{n}|$ = 0 because of the Pauli exclusion between $\spinup$ particles. For the general case see Ref.~\cite{Epelbaum:2009pd,Rupak:2014xza}.  In our three-body system one $\spinup$ and one $\spindown$ fermions are bound together and make a dimer state. We tune the short-range interaction coefficient $C_{0}$ such that the dimer bound state energy is set to $-2.2246$~MeV. This corresponds to a deuteron breakup momentum of about $52.8$ MeV above which the adiabatic Hamiltonian for two clusters is not appropriate as the three-body breakup channels become available. However, the continuum EFT calculations show that the imaginary part of the phase shifts remain quite small even above the breakup momentum. Thus we find that the lattice  phase shift calculations using the two-cluster approximation continues to compare well with the continuum EFT calculation that treat the three-body problem exactly using a Faddeev equation.

The microscopic transfer matrix operator for fermion-dimer system becomes,
\begin{align}
\hat{M}
=
:
\exp\[- \alpha_{t}\,\hat{H}_{0} -\alpha_{t}\, C_{0}\,
\sum_{\vec{n}}
\rho_{\spinup}(\vec{n})
\rho_{\spindown}(\vec{n})
-\alpha_{t}\, 
\alpha_{\text{EM}}\,
Z_{\spinup}^{2}\,
\sum_{\vec{n}}
\frac{\rho_{\spinup}(\vec{n})\rho_{\spinup}(\vec{n})}
{|\vec{n}|} \]
:\,.
\label{eqn:Transfer-matrix-005}
\end{align}
We perform exact lattice calculations to construct the radial adiabatic transfer matrix for the $S$-wave and $P$-wave channels and for various Euclidean projection time steps $L_{t}$. Then we compute the radial adiabatic transfer matrix and use the Schr\"odinger equation to obtain the spectrum to be used in the scattering phase shift calculations as we have discussed in Section~\ref{sec:Scattering-phase-shifts}.

%-----------  Figure ------------------
\begin{figure}[!ht]
\includegraphics[width=0.9\textwidth]{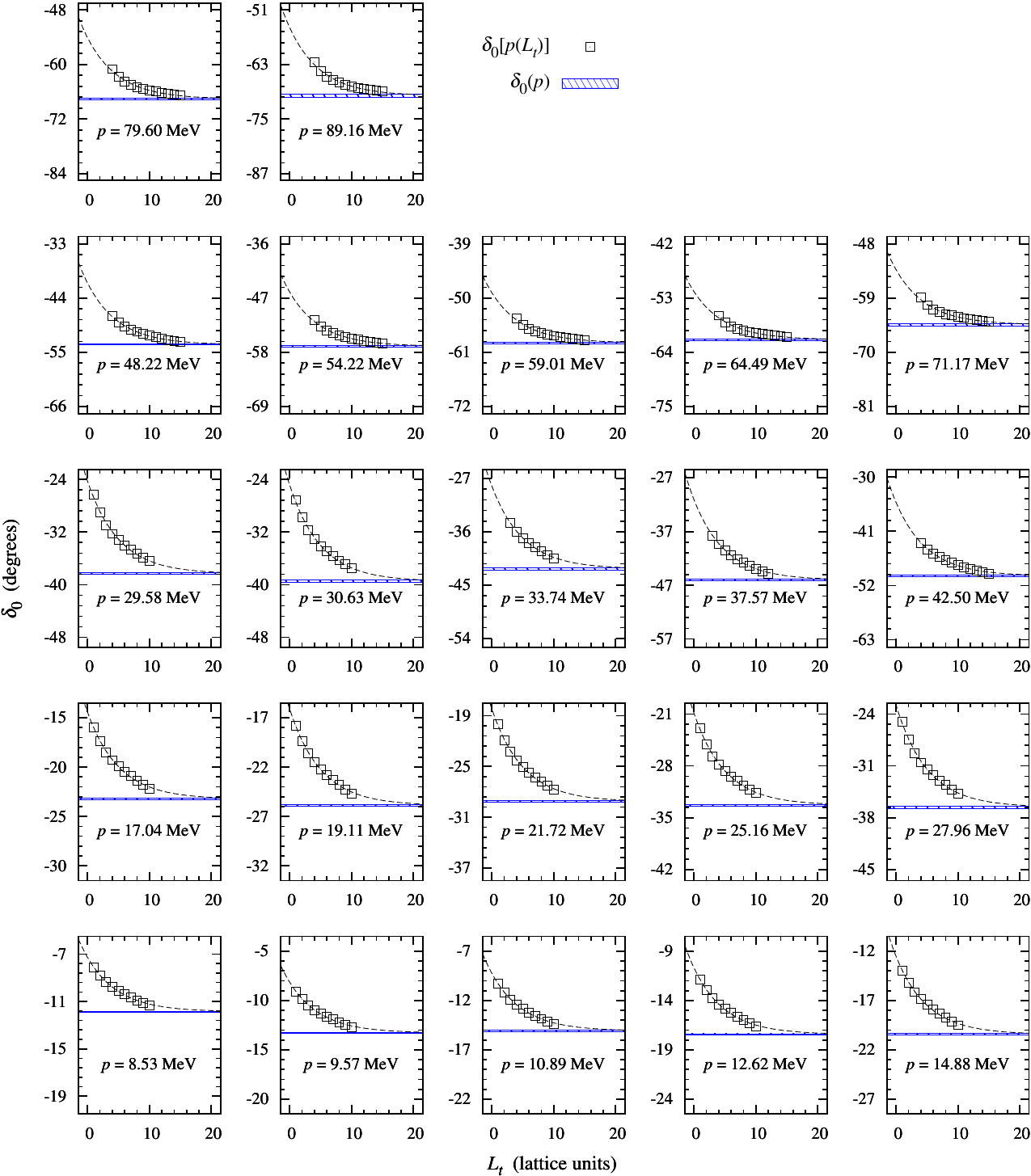}
\caption{Plots of the spin-quartet $S$-wave neutron-deuteron scattering phase shift extrapolation to the limit $L_{t} \to \infty$. The open squares are the the scattering phase shifts versus the Euclidean time steps $L_{t}$ at various relative momenta. The dashed lines are fits to the data. The blue-hatched regions	show the extrapolation with the one standard deviation error estimate.}
\label{fig:Extrapolation-LO-Swave}
\end{figure}
%-----------  Figure ------------------
\begin{figure}[!ht]
\includegraphics[width=0.9\textwidth]{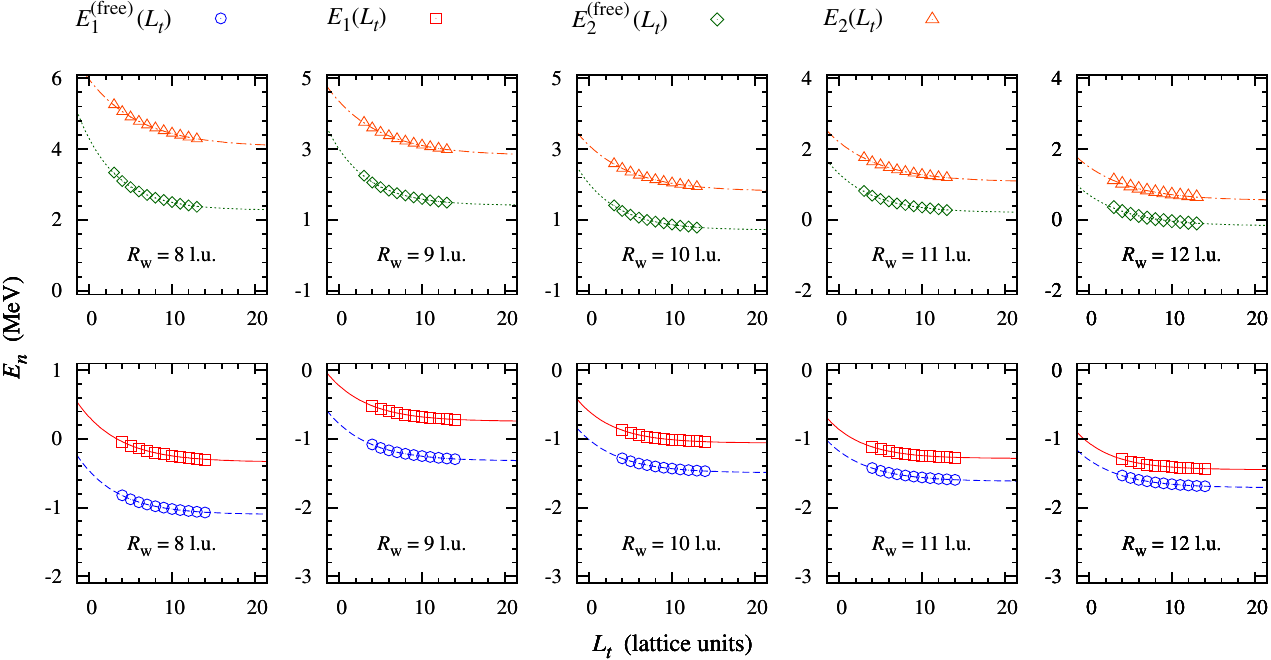}
\caption{Plots of the energy extrapolations to the limit $L_{t} \to \infty$ for the free (trivial) and interacting neutron-deuteron systems in the spin-quartet $S$ channel. The open circles are the first state free energies, the open squares are the first state energies, the open diamonds are the second state free energies, and the open triangles are the second state energies. The lines are fits to the data.}
\label{fig:EnergyExtrapolation-LO-Swave}
\end{figure}
In Fig.~\ref{fig:Extrapolation-LO-Swave} we show the $S$-wave phase shifts in the spin-quartet channel for neutron-deuteron scattering as a function of Euclidean time step $L_{t}$ for various relative momenta. The open squares are the scattering phase shifts, the dashed lines indicate fits of Eq.~(\ref{eqn:phase-shifts-005}) to the lattice data, and the blue-hatched regions show the extrapolations. We also demonstrate the energy extrapolation to the limit $L_{t} \to \infty$ for the trivial and interacting neutron-deuteron system in the spin-quartet-$S$ channel in Fig.~\ref{fig:EnergyExtrapolation-LO-Swave}. We compute the scattering phase shifts using two different radial scattering waves. We refer to these as the first and second states. The open circles (open diamonds) are the first (second) state free neutron-deuteron energies, the open squares (open triangles) are the first (second) state energies, and the lines are fits of Eq.~(\ref{eqn:spectrum-005}) to the  lattice data.

%-----------  Figure ------------------
\begin{figure}[!ht]
\includegraphics[width=0.9\textwidth]{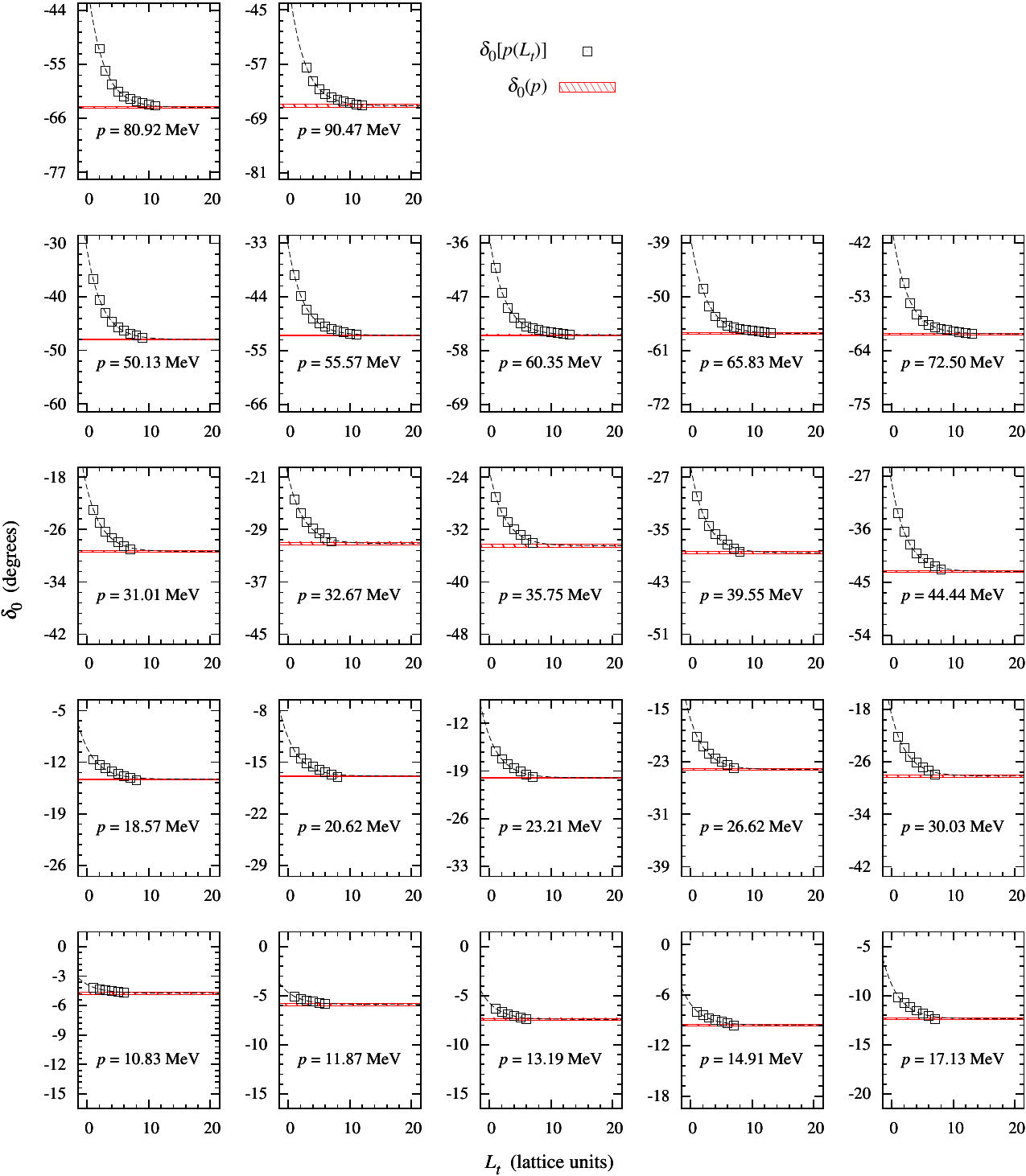}
\caption{Plots of the spin-quartet $S$-wave proton-deuteron scattering phase shift extrapolation to the limit $L_{t} \to \infty$. The open squares are the the scattering phase shifts versus the Euclidean time steps $L_{t}$ at various relative momenta. The dashed lines are fits to the data. The red-hatched regions	show the extrapolation with the one standard deviation error estimate.}
\label{fig:Extrapolation-EM-Swave}
\end{figure}
%-----------  Figure ------------------
\begin{figure}[!ht]
\includegraphics[width=0.9\textwidth]{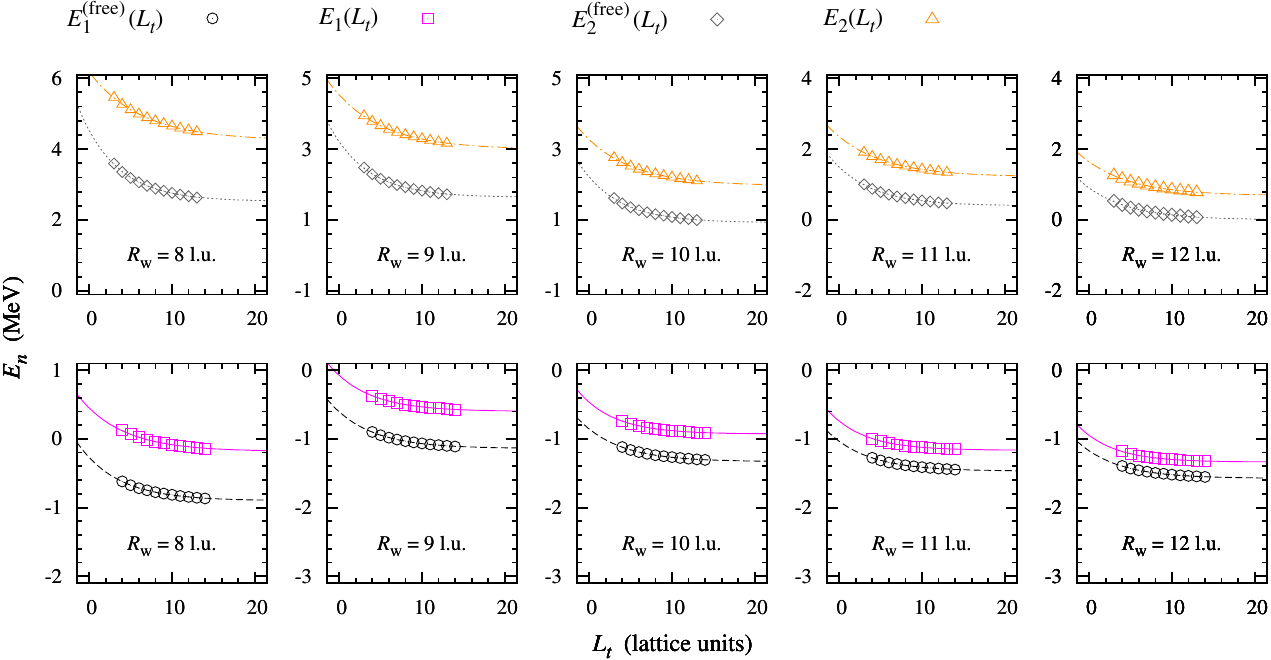}
\caption{Plots of the energy extrapolations to the limit $L_{t} \to \infty$ for the free (trivial) and interacting proton-deuteron systems in the spin-quartet $S$ channel. The open circles are the first state free energies, the open squares are the first state energies, the open diamonds are the second state free energies, and the open triangles are the second state energies. The lines are fits to the data.}
\label{fig:EnergyExtrapolation-EM-Swave}
\end{figure}
%-----------  Figure ------------------
\begin{figure}[!ht]
\centering{
\includegraphics[width=0.6\textwidth]{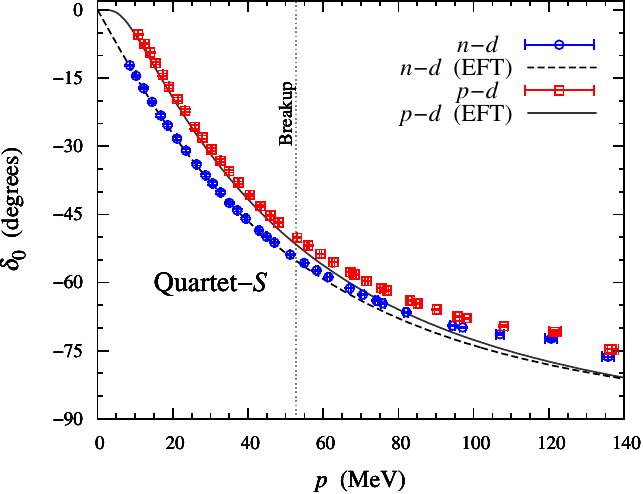}}
\caption{Plots of the S-wave scattering phase shifts versus momentum. The scattering phase shifts are computed from the radial adiabatic transfer matrix constructed by connecting a trivial radial adiabatic transfer matrix from a lattice of size ${82}^{3}\times L_{t}$  and an interacting radial transfer matrix from a lattice of size ${15}^{3}\times L_{t}$  for a fermion-dimer system. The fermion-dimer breakup threshold is about $52.8$~MeV.}
\label{fig:phase-shifts-Swave}
\end{figure}
Similarly, the $S$-wave proton-deuteron scattering phase shifts in the spin-quartet channel versus the Euclidean time steps $L_{t}$ for various relative momenta are shown in Fig.~\ref{fig:Extrapolation-EM-Swave}. The open squares are the scattering phase shifts, the dashed lines are fits of Eq.~(\ref{eqn:phase-shifts-005}) to the data, and the red-hatched regions show the extrapolations. Also the energy extrapolation to the limit $L_{t} \to \infty$ for the trivial and interacting proton-deuteron system in the spin-quartet-$S$ channel is shown in Fig.~\ref{fig:EnergyExtrapolation-EM-Swave}. The open circles (open diamonds) are the first (second) state free proton-deuteron energies, the open squares (open triangles) are the first (second) state energies, and the lines are fits of Eq.~(\ref{eqn:spectrum-005}) to the data.

We compare our lattice results of the $S$-wave scattering phase shifts to the elastic scattering phase shift calculations in the continuum and infinite-volume limits at leading order in pionless effective field theory~\cite{Bedaque:1997qi,Bedaque:1998mb,Gabbiani:1999yv,Bedaque:2002yg,Rupak:2001ci,Konig:2011yq}. 
We emphasize again that we compare the LO lattice EFT data to the same LO continuum EFT results for nucleon-deuteron scattering. Experimental data for this exist~\cite{vanOers:1967,Schmelzbach:1972, Huttel:1983}. So do higher order EFT calculations and potential model calculations that describe the experimental data well, for example Refs.~\cite{Rupak:2001ci,Bedaque:2002yg, Konig:2011yq,Kievsky:1998gt,Deltuva:2005xa} and references there in. However, it is more meaningful to compare the same LO EFT lattice and continuum calculation for benchmarking the accuracy of the numerical methods.

In Fig.~\ref{fig:phase-shifts-Swave} we show the phase shifts versus relative momentum for the $S$-wave in the spin-quartet channel of neutron-deuteron and proton-deuteron scattering. The blue circles and red squares are the lattice results for the neutron-deuteron phase shifts and proton-deuteron phase shifts, respectively. The dashed line indicates the neutron-deuteron scattering phase shifts and the solid line shows the proton-deuteron scattering phase shifts in the continuum and infinite volume limits~\cite{Bedaque:1997qi,Bedaque:1998mb,Gabbiani:1999yv,Bedaque:2002yg,Rupak:2001ci,Konig:2011yq}. We find quite good agreement between the lattice and continuum results. Due to the fact that we use the $\mathcal{O}(a^{4})$-improved lattice action, the agreement in the neutron-proton scattering phase shifts holds for considerably higher momenta than in previous works~\cite{Pine:2013zja,Elhatisari:2014lka,Rokash:2015hra}.

%-----------  Figure ------------------
\begin{figure}[!ht]
\includegraphics[width=0.9\textwidth]{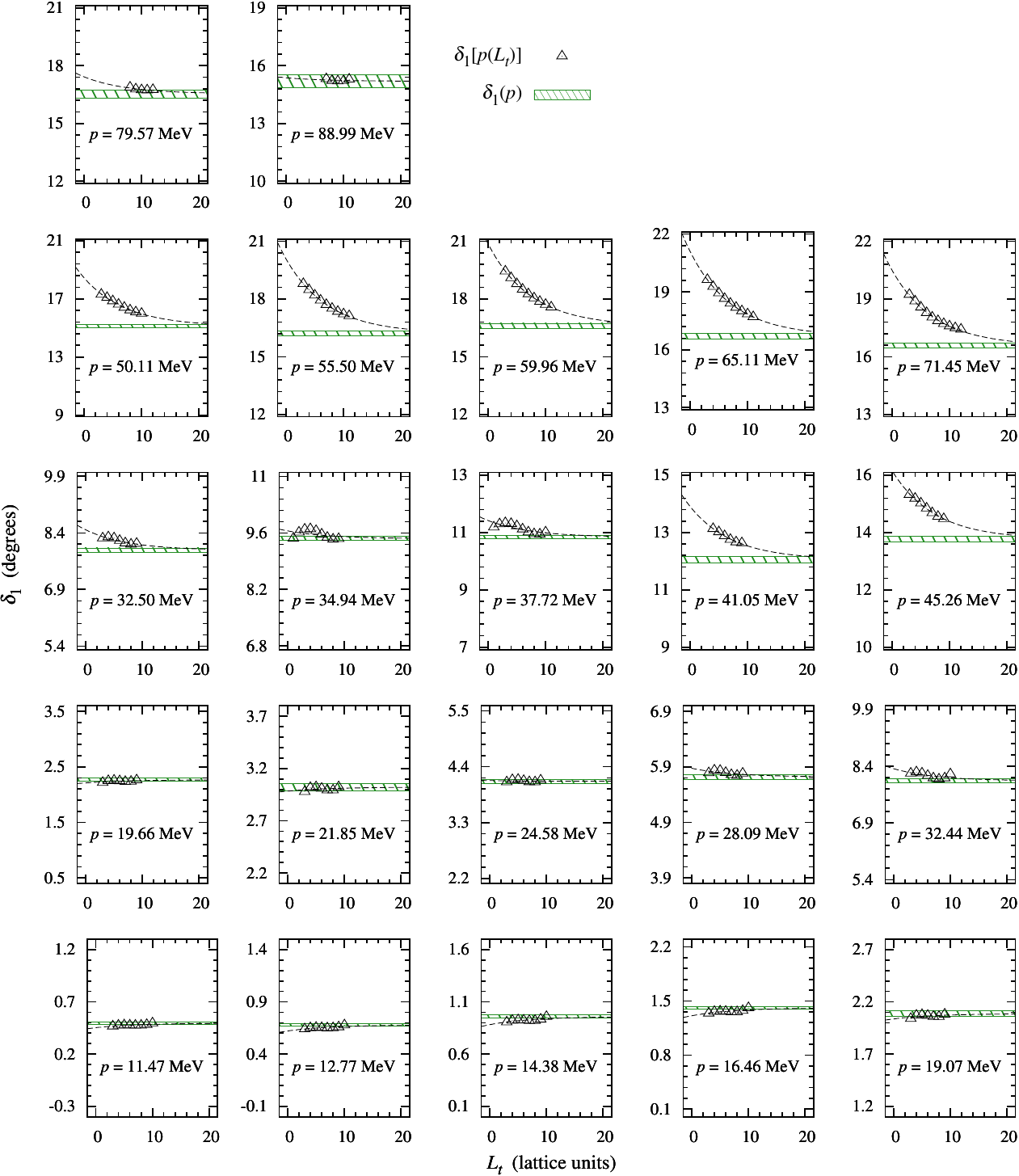}
\caption{Plots of the $P$-wave neutron-deuteron scattering phase shift extrapolation to the limit $L_{t} \to \infty$. The open triangles  are the the scattering phase shifts versus $L_{t}$ at various relative momenta. The dashed lines are fits to the data. The green-hatched regions indicate the one standard deviation error estimate of the extrapolation.}
\label{fig:Extrapolation-LO-Pwave}
\end{figure}
%-----------  Figure ------------------
\begin{figure}[!ht]
\includegraphics[width=0.9\textwidth]{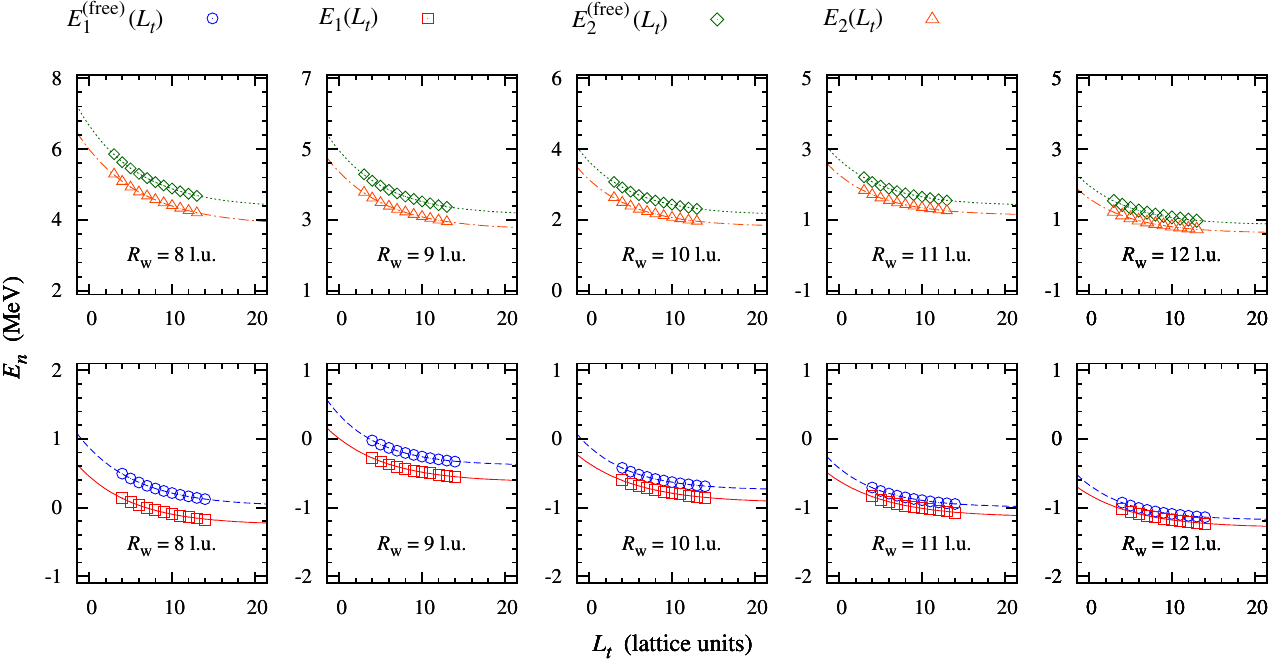}
\caption{Plots of the energy extrapolations to the limit $L_{t} \to \infty$ for the free (trivial) and interacting neutron-deuteron systems in the spin-quartet $P$ channel. The open circles are the first state free energies, the open squares are the first state energies, the open diamonds are the second state free energies, and the open triangles are the second state energies. The lines are fits to the data.}
\label{fig:EnergyExtrapolation-LO-Pwave}
\end{figure}
In Fig.~\ref{fig:Extrapolation-LO-Pwave} we show the $P$-wave phase shifts in the spin-quartet channel of neutron-deuteron scattering versus the Euclidean time steps $L_{t}$ for various relative momenta. The open triangles are the scattering phase shifts, the dashed lines indicate fits of Eq.~(\ref{eqn:phase-shifts-005}) to the data, and the green-hatched regions show the extrapolations with one standard deviation error estimate. Also, the energy extrapolation to the limit $L_{t} \to \infty$ for free and interacting neutron-deuteron in the spin-quartet-$P$ channel are displayed in Fig.~\ref{fig:EnergyExtrapolation-LO-Pwave}. The open circles (open diamonds) are the first (second) state free neutron-deuteron energies, the open squares (open triangles) are the first (second) state energies, and the lines are fits of Eq.~(\ref{eqn:spectrum-005}) to the data.

%-----------  Figure ------------------
\begin{figure}[!ht]
\includegraphics[width=0.9\textwidth]{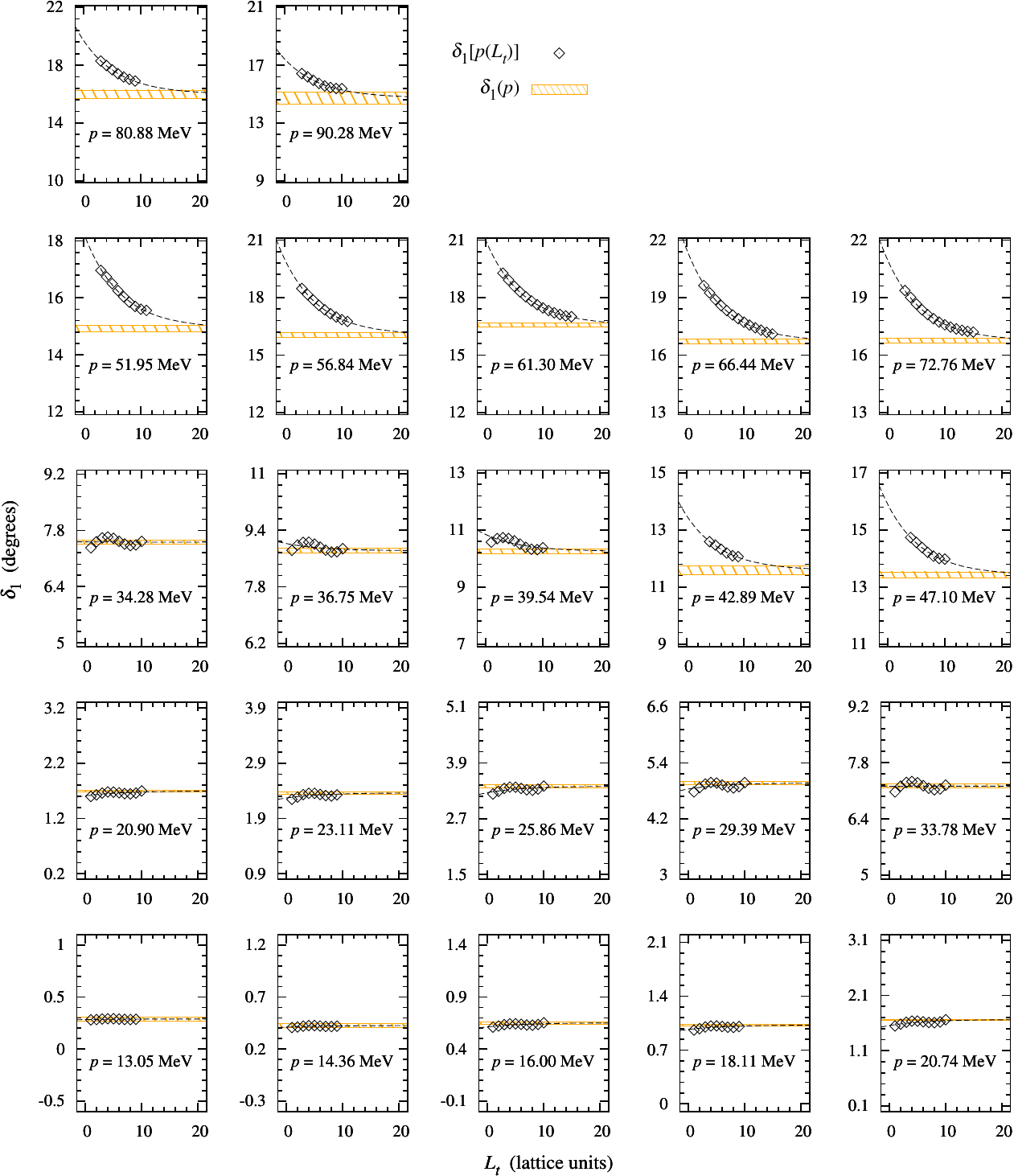}
\caption{Plots of the $P$-wave proton-deuteron scattering phase shift extrapolation to the limit $L_{t} \to \infty$. The open squares are the the scattering phase shifts versus $L_{t}$ at various relative momenta. The dashed lines are fits to the data. The orange-hatched regions	indicate the one standard deviation error estimate of the extrapolation.}
\label{fig:Extrapolation-EM-Pwave}
\end{figure}
%-----------  Figure ------------------
\begin{figure}[!ht]
\includegraphics[width=0.9\textwidth]{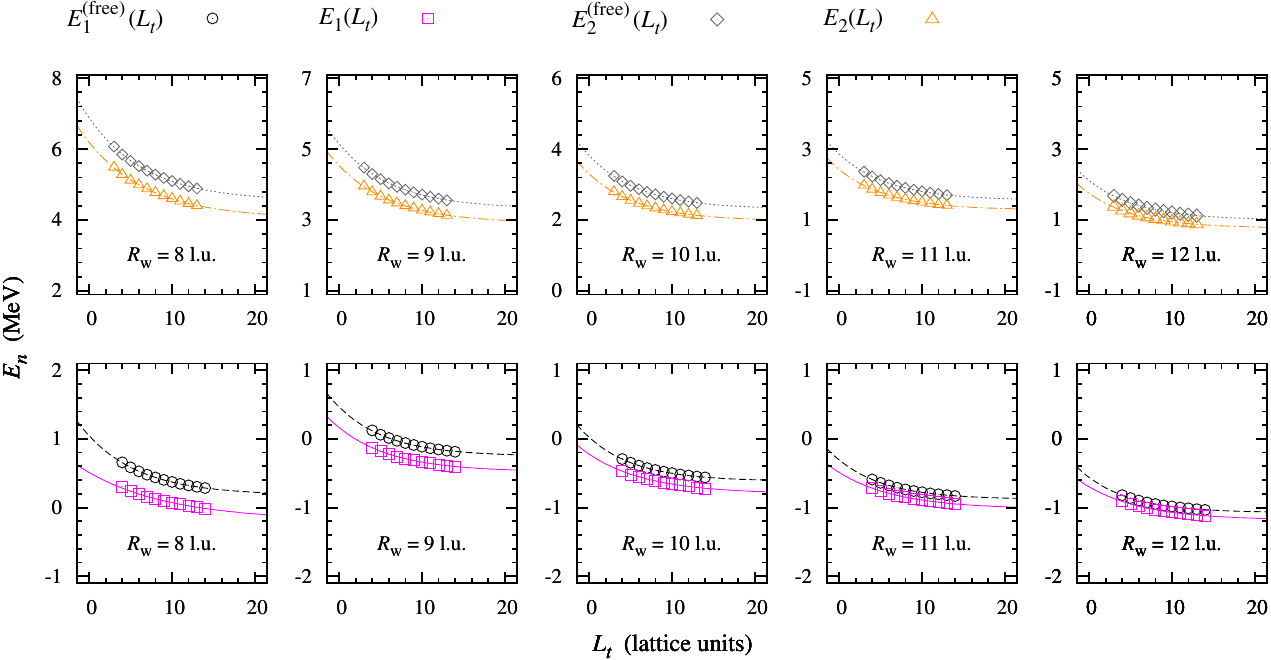}
\caption{Plots of the energy extrapolations to the limit $L_{t} \to \infty$ for the free (trivial) and interacting proton-deuteron systems in the spin-quartet $P$ channel. The open circles are the first state free energies, the open squares are the first state energies, the open diamonds are the second state free energies, and the open triangles are the second state energies. The lines are fits to the data.}
\label{fig:EnergyExtrapolation-EM-Pwave}
\end{figure}
The $P$-wave phase shifts in the spin-quartet channel of  proton-deuteron scattering versus the Euclidean time steps $L_{t}$ for various relative momenta are shown in Fig.~\ref{fig:Extrapolation-EM-Pwave}. The open diamonds are the scattering phase shifts, the dashed lines are fits of Eq.~(\ref{eqn:phase-shifts-005}) to the data, and the orange-hatched regions indicate the extrapolations The energy extrapolation to the limit $L_{t} \to \infty$ for free and interacting proton-deuteron in the spin-quartet-$P$ channel is displayed in Fig.~\ref{fig:EnergyExtrapolation-EM-Pwave}. The open circles (open diamonds) are the first (second) state free proton-deuteron energies, the open squares (open triangles) are the first (second) state energies, and the lines are fits of Eq.~(\ref{eqn:spectrum-005}) to the data.

%-----------  Figure ------------------
\begin{figure}[!ht]
\centering{
\includegraphics[width=0.6\textwidth]{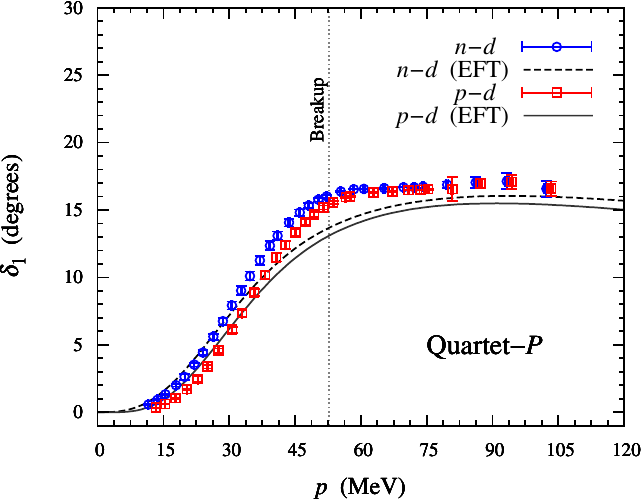}}
\caption{Plots of the P-wave scattering phase shifts versus momentum. The scattering phase shifts are computed from the radial adiabatic transfer matrix constructed by connecting a trivial radial adiabatic transfer matrix from a lattice of size ${82}^{3}\times L_{t}$  and an interacting radial transfer matrix from a lattice of size ${15}^{3}\times L_{t}$  for a fermion-dimer system. The fermion-dimer breakup threshold is about $52.8$~MeV.}
\label{fig:phase-shifts-Pwave}
\end{figure}
For nucleon-deuteron scattering we compare our lattice calculations of the $P$-wave phase shifts  to the elastic scattering phase shift calculations in the continuum and infinite-volume limits at leading order in pionless effective field theory. In Fig.~\ref{fig:phase-shifts-Pwave}, we show the $P$-wave phase shifts for neutron-deuteron and proton-deuteron scattering in the spin quartet channel. The blue circles are the lattice results that we compare with the corresponding results from EFT in the continuum and infinite volume limits shown as the dashed line~\cite{Bedaque:1997qi,Bedaque:1998mb,Gabbiani:1999yv,Bedaque:2002yg,Vanasse:2013sda}. Similarly, the red squares compare the lattice results for proton-deuteron scattering with the corresponding  EFT calculations in the continuum and infinite volume limits represented by the solid line~\cite{Rupak:2001ci,Konig:2011yq,Vanasse:2014kxa}. We find good agreement between the lattice and continuum results for neutron-deuteron  and proton-deuteron scattering.  The deviation with the continuum results at larger momenta is due to nonzero lattice spacing effects (observed in previous work [12]), and also due to the fact that we are above the threshold for deuteron breakup.

%===============================  Summary ======================
\section{Summary and Discussion}
\label{sec:summary-and-discussion}
%================================================================

In this paper we have presented several new methods for studying two-cluster scattering on the lattice using the adiabatic projection method. As extensively discussed in earlier pioneering works~\cite{Pine:2013zja,Elhatisari:2014lka,Rokash:2015hra}, the adiabatic projection method is a general framework for studying scattering and reactions of clusters on the lattice. The method uses the initial cluster states and the Euclidean time projection to construct a low energy effective theory for the participating clusters, and in the large Euclidean projection time limit the method gives the exact descriptions of the low-lying states for two clusters. In this study we use a new method which parametrizes the initial cluster states in spherical coordinates by projecting the initial cluster states onto spherical harmonics. Furthermore, we have also introduced a radial bin size $a_{R}$, and we have binned the initial cluster states by grouping the lattice points with radial distance into a given band of width $a_{R}$. 

We found that the new methods provide crucial improvements in the efficiency of the two-cluster adiabatic transfer matrix calculations without introducing additional systematic errors into the calculations. The improvements were summarized in Table~\ref{table:comp-scaling} in terms of the dimensions of the adiabatic transfer matrix for two-cluster systems. In order to check the systematic errors due to projecting the initial cluster states onto spherical harmonics and defining the radial bin size $a_{R}$, we benchmarked these methods using two-particle scattering on the lattice. We found that the corresponding systematic errors are numerically very small, at most a couple percent of relative error in the phase shift. 

We employed the new methods discussed in this paper to study a two-cluster system using the adiabatic projection method. We considered fermion-dimer scattering with and without a long-range Coulomb potential, which correspond to proton-deuteron and neutron-deuteron scattering in the spin-quartet channel, respectively. For the analyses of the scattering phase shifts we used the spherical wall method. We computed the $S$-wave and $P$-wave phase shifts for neutron-deuteron and proton-deuteron scattering in the spin-quartet channel, and we compared our results to the available calculations for the corresponding systems in the continuum and the infinite volume limits. We found that our lattice results are in good agreements with the continuum calculations.

The methods that we have presented in this paper are developed and designed to improve the efficiency of large-scale calculations of nucleus-nucleus scattering and reactions using Monte Carlo calculations. Recently, these methods are employed in the calculation of the $S$-wave alpha-alpha scattering phase shifts using the adiabatic projection and the calculations are greatly improved~\cite{Elhatisari:2015iga,Elhatisari:2016owd}.

\acknowledgements
We acknowledge the supports from the DFG (CRC 110), U.S. Department of Energy grant DE-FG02-03ER41260 and U.S. National Science Foundation grant PHY-1307453. The work of UGM is also supported in part by the Chinese Academy of Sciences (CAS) President's 
International Fellowship Initiative (PIFI) (Grant No. 2015VMA076). 

\appendix

\section{Extrapolation}
\label{sec:extrapolation}

We discuss the Euclidean time extrapolations to eliminate the contaminations from excited states to the energy and the scattering phase shifts. Let us consider Euclidean time projection with a single initial cluster state which has a nonzero overlap with the ground state,
\begin{align} 
\ket{\psi} = \sum_{i} \, c_{i}\, \ket{\phi_{i}}\,,
\label{eqn:eigenstates-001}
\end{align}
where $ \ket{\phi_{i}}$ are the energy eigenstates with energies $E_{i}$ in order of increasing energy
\begin{align} 
E_{0} \leq E_{1} \leq E_{2} \ldots
\label{eqn:eigenenergies-001}
\end{align}
The energy estimate at $L_{t}$ Euclidean time step is 
\begin{align}
E(L_{t}) 
&
=  -\frac{1}{\alpha_{t}}
\log\[
\frac{\prescript{}{n_{t}}{\braket{\psi|\hat{M}|\psi}}_{n_{t}}^{}}
{\prescript{}{n_{t}}{\braket{\psi|\psi}}_{n_{t}}^{}}
\]
\,,
\label{eqn:energy-001}
\end{align}
or explicitly~\citep{Lee:2008xsa,Katterjohn:xxx},
\begin{align}
E(L_{t}) 
=
\frac{1}{\alpha_{t}}
\log\[
\frac{\sum_{i}\,c_{i}^{2}\,e^{-E_{i}\,(L_{t}-1) \alpha_{t}}}
{\sum_{i}\,c_{i}^{2}\,e^{-E_{i}\,L_{t}\,\alpha_{t}}}
\]
\,.
\label{eqn:energy-005}
\end{align}
Eq.~(\ref{eqn:energy-005}) can be rewritten in terms of the ground state energy and the contributions from the excited states as,
\begin{align}
E(L_{t}) 
%&
%=
%\frac{1}{a_{t}}
%\log\[
%e^{E_{0}\,a_{t}}
%\,
%\frac{1+ \sum_{i>0}\,(\frac{c_{i}}{c_{0}})^{2}\,e^{-(E_{i}-E_{0})(L_{t}-1)\alpha_{t}}}
%{1 + \sum_{i>0}\,(\frac{c_{i}}{c_{0}})^{2}\,e^{-(E_{i}-E_{0})\,L_{t}\alpha_{t}}}
%\]
%\,,
%\\
%&
=
E_{0} + 
\frac{1}{\alpha_{t}}
\log\[
\frac{1+ \sum_{i>0}\,(\frac{c_{i}}{c_{0}})^{2}\,e^{-(E_{i}-E_{0})(L_{t}-1)\alpha_{t}}}
{1 + \sum_{i>0}\,(\frac{c_{i}}{c_{0}})^{2}\,e^{-(E_{i}-E_{0})\,L_{t}\alpha_{t}}}
\]
\,.
\label{eqn:energy-009}
\end{align}
Then in the large Euclidean time limit and using the fact that $E_{i}-E_{0} \ll \alpha_{t}^{-1}$ for low-energy excitations, Eq.~(\ref{eqn:energy-009}) can be further simplified to
\begin{align}
E(L_{t}) 
\cong
E_{0} + 
\sum_{i>0}
\,(E_{i}-E_{0})
\(\frac{c_{i}}{c_{0}}\)^{2}
\,e^{-(E_{i}-E_{0})\,L_{t}\alpha_{t}}
\,.
\label{eqn:energy-013}
\end{align}

Now we want to find an expression for the scattering phase shift extrapolation. Let us write the scattering phase shift estimate at $L_{t}$ time steps starting from the effective range expansion,
\begin{align}
\delta_{\ell}[p(L_{t})] = \cot^{-1}
\[
p({L_{t}})^{-2\ell-1} \(-\frac{1}{a_{\ell}}+\frac{1}{2}r_{\ell}\,p({L_{t}})^{2}  + \mathcal{O}(p({L_{t}})^{4}) \)
\]
\,,
\label{eqn:phaseshifts-001}
\end{align}
where $a_{\ell}$ and $r_{\ell}$ are the scattering parameters at the two lowest order of the expansion. $p({L_{t}})$ is the momentum estimate at $L_{t}$ time steps and defined in terms of the energy estimate at $L_{t}$ time step given in Eq.~(\ref{eqn:energy-013}),
\begin{align}
p({L_{t}}) = \sqrt{{2\mu\, E(L_{t})}}= \sqrt{p^{2} + A(L_{t})}
\label{eqn:momentum-001}
\end{align}
where $p$ is the momentum in the large $L_{t}$ limit and $A(L_{t})$ is the residual correction at $L_{t}$,
\begin{align}
A(L_{t})
\cong
\sum_{i>0}
\,(p^{2}_{i}-p^{2})
\(\frac{c_{i}}{c_{0}}\)^{2}
\,e^{-(E_{i}-E_{0})\,L_{t}\alpha_{t}}\,.
\end{align}
Now we insert Eq.~(\ref{eqn:momentum-001}) into Eq.~(\ref{eqn:phaseshifts-001}), then after some manipulations, we obtain the following expression for the scattering phase shift estimate at $L_{t}$ time steps,
\begin{align}
\delta_{\ell}[p(L_{t})] = 
\delta_{\ell}(p) 
+
c_{\ell}\, \frac{A(L_{t})}{p^{2}} + \mathcal{O}(A(L_{t})^{2})\,,
\label{eqn:phaseshifts-009}
\end{align}
where $c_{\ell}$ absorbs the contaminations from the scattering phase shifts for excited states. For the system where the splittings from excited states are large, this expression can be further simplified to Eq.~(\ref{eqn:phase-shifts-005}).

\newpage

%\bibliographystyle{apsrev}
%\bibliography{references}

\newpage

\begin{thebibliography}{45}
	
	\expandafter\ifx\csname natexlab\endcsname\relax\def\natexlab#1{#1}\fi
	\expandafter\ifx\csname bibnamefont\endcsname\relax
	\def\bibnamefont#1{#1}\fi
	\expandafter\ifx\csname bibfnamefont\endcsname\relax
	\def\bibfnamefont#1{#1}\fi
	\expandafter\ifx\csname citenamefont\endcsname\relax
	\def\citenamefont#1{#1}\fi
	\expandafter\ifx\csname url\endcsname\relax
	\def\url#1{\texttt{#1}}\fi
	\expandafter\ifx\csname urlprefix\endcsname\relax\def\urlprefix{URL }\fi
	\providecommand{\bibinfo}[2]{#2}
	\renewcommand{\eprint}[2][]{[#2]}
	
	
	%\cite{Nollett:2006su}
	\bibitem{Nollett:2006su} 
	K.~M.~Nollett, S.~C.~Pieper, R.~B.~Wiringa, J.~Carlson and G.~M.~Hale,
	%``Quantum Monte Carlo calculations of neutron-alpha scattering,''
	Phys.\ Rev.\ Lett.\  {\bf 99}, 022502 (2007)
	%doi:10.1103/PhysRevLett.99.022502
	[nucl-th/0612035].
	%%CITATION = doi:10.1103/PhysRevLett.99.022502;%%
	
	
	%\cite{Quaglioni:2008sm}
	\bibitem{Quaglioni:2008sm} 
	S.~Quaglioni and P.~Navratil,
	%``Ab Initio Many-Body Calculations of n-H-3, n-He-4, p-He-3,4,and and n-Be-10 Scattering,''
	Phys.\ Rev.\ Lett.\  {\bf 101}, 092501 (2008)
	%doi:10.1103/PhysRevLett.101.092501
	[arXiv:0804.1560 [nucl-th]].
	%%CITATION = doi:10.1103/PhysRevLett.101.092501;%%
	
	
	%\cite{Navratil:2010jn}
	\bibitem{Navratil:2010jn} 
	P.~Navratil, R.~Roth and S.~Quaglioni,
	%``Ab initio many-body calculations of nucleon scattering on $^4He, ^7Li, ^7Be, ^{12}C and ^{16}$O,''
	Phys.\ Rev.\ C {\bf 82}, 034609 (2010)
	%doi:10.1103/PhysRevC.82.034609
	[arXiv:1007.0525 [nucl-th]].
	%%CITATION = doi:10.1103/PhysRevC.82.034609;%%
	
	
	%\cite{Navratil:2011sa}
	\bibitem{Navratil:2011sa} 
	P.~Navratil, R.~Roth and S.~Quaglioni,
	%``Ab initio many-body calculation of the $^{7}$Be$(p,\gamma)^8$B radiative capture,''
	Phys.\ Lett.\ B {\bf 704}, 379 (2011)
	%doi:10.1016/j.physletb.2011.09.079
	[arXiv:1105.5977 [nucl-th]].
	%%CITATION = doi:10.1016/j.physletb.2011.09.079;%%
	
	
	
	%\cite{Hupin:2013wsa}
	\bibitem{Hupin:2013wsa} 
	G.~Hupin, J.~Langhammer, P.~Navrátil, S.~Quaglioni, A.~Calci and R.~Roth,
	%``Ab initio many-body calculations of nucleon-4He scattering with three-nucleon forces,''
	Phys.\ Rev.\ C {\bf 88}, no. 5, 054622 (2013)
%	doi:10.1103/PhysRevC.88.054622
	[arXiv:1308.2700 [nucl-th]].
	%%CITATION = doi:10.1103/PhysRevC.88.054622;%%
	
	
	%\cite{Hagen:2012rq}
	\bibitem{Hagen:2012rq} 
	G.~Hagen and N.~Michel,
	%``Elastic proton scattering of medium mass nuclei from coupled-cluster theory,''
	Phys.\ Rev.\ C {\bf 86}, 021602 (2012)
%	doi:10.1103/PhysRevC.86.021602
	[arXiv:1206.2336 [nucl-th]].
	%%CITATION = doi:10.1103/PhysRevC.86.021602;%%
	
	
	%\cite{Navratil:2011ay}
	\bibitem{Navratil:2011ay} 
	P.~Navratil and S.~Quaglioni,
	%``Ab initio many-body calculations of deuteron-4He scattering and 6Li states,''
	Phys.\ Rev.\ C {\bf 83}, 044609 (2011)
%	doi:10.1103/PhysRevC.83.044609
	[arXiv:1102.2042 [nucl-th]].
	%%CITATION = doi:10.1103/PhysRevC.83.044609;%%
	
	
	%\cite{Navratil:2011zs}
	\bibitem{Navratil:2011zs} 
	P.~Navratil and S.~Quaglioni,
	%``Ab initio many-body calculations of the 3H(d,n)4He and 3He(d,p)4He fusion,''
	Phys.\ Rev.\ Lett.\  {\bf 108}, 042503 (2012)
%	doi:10.1103/PhysRevLett.108.042503
	[arXiv:1110.0460 [nucl-th]].
	%%CITATION = doi:10.1103/PhysRevLett.108.042503;%%
	
	
	%\cite{Elhatisari:2015iga}
	\bibitem{Elhatisari:2015iga} 
	S.~Elhatisari, D.~Lee, G.~Rupak, E.~Epelbaum, H.~Krebs, T.~A.~Lähde, T.~Luu and U.-G.~Meißner,
	%``Ab initio alpha-alpha scattering,''
	Nature {\bf 528}, 111 (2015)
%	doi:10.1038/nature16067
	[arXiv:1506.03513 [nucl-th]].
	%%CITATION = doi:10.1038/nature16067;%%
	
	
	%\cite{Elhatisari:2016owd}
	\bibitem{Elhatisari:2016owd} 
%	S.~Elhatisari {\it et al.}
	S.~Elhatisari, N.~Li, A.~Rokash, J.~M.~Alarcon, D.~Du, N.~Klein 
	B.~N.~Lu, U.-G.~Mei{\ss}ner, E.~Epelbaum,
	H.~Krebs, T.~A.~L{\"a}hde, D.~Lee and G.~Rupak,
	%``Nuclear binding near a quantum phase transition,''
	arXiv:1602.04539 [nucl-th].
	%%CITATION = ARXIV:1602.04539;%%
	
	
	%\cite{Pine:2013zja}
	\bibitem{Pine:2013zja} 
	M.~Pine, D.~Lee and G.~Rupak,
	%``Adiabatic projection method for scattering and reactions on the lattice,''
	Eur.\ Phys.\ J.\ A {\bf 49}, 151 (2013)
%	doi:10.1140/epja/i2013-13151-3
	[arXiv:1309.2616 [nucl-th]].
	%%CITATION = doi:10.1140/epja/i2013-13151-3;%%
	
	
	%\cite{Elhatisari:2014lka}
	\bibitem{Elhatisari:2014lka} 
	S.~Elhatisari and D.~Lee,
	%``Fermion-dimer scattering using an impurity lattice Monte Carlo approach and the adiabatic projection method,''
	Phys.\ Rev.\ C {\bf 90}, no. 6, 064001 (2014)
%	doi:10.1103/PhysRevC.90.064001
	[arXiv:1407.2784 [nucl-th]].
	%%CITATION = doi:10.1103/PhysRevC.90.064001;%%
	
	
	%\cite{Rokash:2015hra}
	\bibitem{Rokash:2015hra} 
	A.~Rokash, M.~Pine, S.~Elhatisari, D.~Lee, E.~Epelbaum and H.~Krebs,
	%``Scattering cluster wave functions on the lattice using the adiabatic projection method,''
	Phys.\ Rev.\ C {\bf 92}, no. 5, 054612 (2015)
%	doi:10.1103/PhysRevC.92.054612
	[arXiv:1505.02967 [nucl-th]].
	%%CITATION = doi:10.1103/PhysRevC.92.054612;%%
	
	%\cite{Rupak:2013aue}
	\bibitem{Rupak:2013aue} 
	G.~Rupak and D.~Lee,
	%``Radiative capture reactions in lattice effective field theory,''
	Phys.\ Rev.\ Lett.\  {\bf 111}, no. 3, 032502 (2013)
%	doi:10.1103/PhysRevLett.111.032502
	[arXiv:1302.4158 [nucl-th]].
	%%CITATION = doi:10.1103/PhysRevLett.111.032502;%%
	
	
	%\cite{Rupak:2014xza}
	\bibitem{Rupak:2014xza} 
	G.~Rupak and P.~Ravi,
	%``Proton–proton fusion in lattice effective field theory,''
	Phys.\ Lett.\ B {\bf 741}, 301 (2015)
%	doi:10.1016/j.physletb.2014.12.055
	[arXiv:1411.2436 [nucl-th]].
	%%CITATION = doi:10.1016/j.physletb.2014.12.055;%%
	
	
	%\cite{Bedaque:1997qi}
	\bibitem{Bedaque:1997qi} 
	P.~F.~Bedaque and U.~van Kolck,
	%``Nucleon deuteron scattering from an effective field theory,''
	Phys.\ Lett.\ B {\bf 428}, 221 (1998)
%	doi:10.1016/S0370-2693(98)00430-4
	[nucl-th/9710073].
	%%CITATION = doi:10.1016/S0370-2693(98)00430-4;%%
	
	
	%\cite{Chen:1999tn}
	\bibitem{Chen:1999tn} 
	J.~W.~Chen, G.~Rupak and M.~J.~Savage,
	%``Nucleon-nucleon effective field theory without pions,''
	Nucl.\ Phys.\ A {\bf 653}, 386 (1999)
%	doi:10.1016/S0375-9474(99)00298-5
	[nucl-th/9902056].
	%%CITATION = doi:10.1016/S0375-9474(99)00298-5;%%
	
	
	%\cite{Lu:2015riz}
	\bibitem{Lu:2015riz} 
	B.~N.~Lu, T.~A.~Lähde, D.~Lee and U.-G.~Meißner,
	%``Precise determination of lattice phase shifts and mixing angles,''
	arXiv:1506.05652 [nucl-th].
	%%CITATION = ARXIV:1506.05652;%%
	
	
	%\cite{Carlson:1984zz}
	\bibitem{Carlson:1984zz} 
	J.~Carlson, V.~R.~Pandharipande and R.~B.~Wiringa,
	%``Variational calculations of resonant states in $^4$He,''
	Nucl.\ Phys.\ A {\bf 424}, 47 (1984).
%	doi:10.1016/0375-9474(84)90127-1
	%%CITATION = doi:10.1016/0375-9474(84)90127-1;%%
	
	
	%\cite{Borasoy:2007vy}
	\bibitem{Borasoy:2007vy} 
	B.~Borasoy, E.~Epelbaum, H.~Krebs, D.~Lee and U.-G.~Mei{\ss}ner,
	%``Two-particle scattering on the lattice: Phase shifts, spin-orbit coupling, and mixing angles,''
	Eur.\ Phys.\ J.\ A {\bf 34}, 185 (2007)
%	doi:10.1140/epja/i2007-10500-9
	[arXiv:0708.1780 [nucl-th]].
	%%CITATION = doi:10.1140/epja/i2007-10500-9;%%
	
	
	\bibitem{Abramowitz}
	M.~Abramowitz and I.~A.~Stegun,
	\textit{Handbook of Mathematical Functions}, 
	(Dover Publications,
	1970,
	New York)
	
	
	%\cite{Koenig:2012bv}
	\bibitem{Koenig:2012bv} 
	S.~König, D.~Lee and H.~W.~Hammer,
	%``Causality constraints for charged particles,''
	J.\ Phys.\ G {\bf 40}, 045106 (2013)
%	doi:10.1088/0954-3899/40/4/045106
	[arXiv:1210.8304 [nucl-th]].
	%%CITATION = doi:10.1088/0954-3899/40/4/045106;%%
	
	%\cite{Epelbaum:2009pd}
	\bibitem{Epelbaum:2009pd} 
	E.~Epelbaum, H.~Krebs, D.~Lee and U.-G.~Mei{\ss}ner,
	%``Lattice effective field theory calculations for A = 3,4,6,12 nuclei,''
	Phys.\ Rev.\ Lett.\  {\bf 104}, 142501 (2010)
%	doi:10.1103/PhysRevLett.104.142501
	[arXiv:0912.4195 [nucl-th]].
	%%CITATION = doi:10.1103/PhysRevLett.104.142501;%%
	
	%\cite{Gabbiani:1999yv}
	\bibitem{Gabbiani:1999yv} 
	F.~Gabbiani, P.~F.~Bedaque and H.~W.~Grie{\ss}hammer,
	%``Higher partial waves in an effective field theory approach to nd scattering,''
	Nucl.\ Phys.\ A {\bf 675}, 601 (2000)
%	doi:10.1016/S0375-9474(00)00181-0
	[nucl-th/9911034].
	%%CITATION = doi:10.1016/S0375-9474(00)00181-0;%%
	
	
	%\cite{Konig:2011yq}
	\bibitem{Konig:2011yq} 
	S.~K\"onig and H.~W.~Hammer,
	%``Low-energy p-d scattering and He-3 in pionless EFT,''
	Phys.\ Rev.\ C {\bf 83}, 064001 (2011)
%	doi:10.1103/PhysRevC.83.064001
	[arXiv:1101.5939 [nucl-th]].
	%%CITATION = doi:10.1103/PhysRevC.83.064001;%%
	
	%\cite{Bedaque:1998mb}
	\bibitem{Bedaque:1998mb} 
	P.~F.~Bedaque, H.~W.~Hammer and U.~van Kolck,
	%``Effective theory for neutron deuteron scattering: Energy dependence,''
	Phys.\ Rev.\ C {\bf 58}, 641 (1998)
%	doi:10.1103/PhysRevC.58.641, 10.1103/PhysRevC.58.R641
	[nucl-th/9802057].
	%%CITATION = doi:10.1103/PhysRevC.58.641, 10.1103/PhysRevC.58.R641;%%
	
	%\cite{Bedaque:2002yg}
	\bibitem{Bedaque:2002yg} 
	P.~F.~Bedaque, G.~Rupak, H.~W.~Grie{\ss}hammer and H.~W.~Hammer,
	%``Low-energy expansion in the three-body system to all orders and the triton channel,''
	Nucl.\ Phys.\ A {\bf 714}, 589 (2003)
%	doi:10.1016/S0375-9474(02)01402-1
	[nucl-th/0207034].
	%%CITATION = doi:10.1016/S0375-9474(02)01402-1;%%
	
	
	%\cite{Rupak:2001ci}
	\bibitem{Rupak:2001ci} 
	G.~Rupak and X.~W.~Kong,
	%``Quartet S wave p d scattering in EFT,''
	Nucl.\ Phys.\ A {\bf 717}, 73 (2003)
%	doi:10.1016/S0375-9474(03)00638-9
	[nucl-th/0108059].
	%%CITATION = doi:10.1016/S0375-9474(03)00638-9;%%
	
	
	\bibitem{vanOers:1967}
	W.~T.~H.~van~Oers and J.~D.~Segrave,
	Phys.\ Lett.\ B {\bf 24}, 562 (1967)


	\bibitem{Schmelzbach:1972}
	P.~A.~Schmelzbach, W.~Gr{\"u}bler, R.~E.~White, V.~K{\"o}nig,
	R.~Risler and P.~Marmier,
	Nucl.\ Phys.\ A {\bf 197}, 273 (1972)

	\bibitem{Huttel:1983}
	E.~Huttel, W.~Arnold, H.~Baumgart, H.~Berg and G.~Clausnitzer,
	Nucl.\ Phys.\ A {\bf 406}, 443 (1983)
	
	
	%\cite{Kievsky:1998gt}
	\bibitem{Kievsky:1998gt} 
	A.~Kievsky, M.~Viviani, S.~Rosati, D.~H{\"u}ber, W.~Gloeckle, H.~Kamada, H.~Witala and J.~Golak,
	%``Benchmark calculations for polarization observables in three nucleon scattering,''
	Phys.\ Rev.\ C {\bf 58}, 3085 (1998)
%	doi:10.1103/PhysRevC.58.3085
	[nucl-th/9807061].
	%%CITATION = doi:10.1103/PhysRevC.58.3085;%%
	
	
	
	%\cite{Deltuva:2005xa}
	\bibitem{Deltuva:2005xa} 
	A.~Deltuva, A.~C.~Fonseca, A.~Kievsky, S.~Rosati, P.~U.~Sauer and M.~Viviani,
	%``Benchmark calculation for proton-deuteron elastic scattering observables including Coulomb,''
	Phys.\ Rev.\ C {\bf 71}, 064003 (2005)
%	doi:10.1103/PhysRevC.71.064003
	[nucl-th/0503015].
	%%CITATION = doi:10.1103/PhysRevC.71.064003;%%
	
	
	%\cite{Lee:2008xsa}
	\bibitem{Lee:2008xsa} 
	D.~Lee,
	%``The Ground state energy at unitarity,''
	Phys.\ Rev.\ C {\bf 78}, 024001 (2008)
%	doi:10.1103/PhysRevC.78.024001
	[arXiv:0803.1280 [nucl-th]].
	%%CITATION = doi:10.1103/PhysRevC.78.024001;%%
	
	
	
	\bibitem{Katterjohn:xxx}
	K.~Katterjohn, S.~Elhatisari, D.~Lee,
	U.-G.~Mei{\ss}ner and G.~Rupak,
	work in progress.

	%\cite{Luscher:1986pf}
	\bibitem{Luscher:1986pf} 
	M.~L\"uscher,
	%``Volume Dependence of the Energy Spectrum in Massive Quantum Field Theories. 2. Scattering States,''
	Commun.\ Math.\ Phys.\  {\bf 105}, 153 (1986).
%	doi:10.1007/BF01211097
	%%CITATION = doi:10.1007/BF01211097;%%
	
	
	%\cite{Luscher:1990ux}
	\bibitem{Luscher:1990ux} 
	M.~L\"uscher,
	%``Two particle states on a torus and their relation to the scattering matrix,''
	Nucl.\ Phys.\ B {\bf 354}, 531 (1991).
%	doi:10.1016/0550-3213(91)90366-6
	%%CITATION = doi:10.1016/0550-3213(91)90366-6;%%
	
	
	%\cite{Vanasse:2014kxa}
	\bibitem{Vanasse:2014kxa} 
	J.~Vanasse, D.~A.~Egolf, J.~Kerin, S.~König and R.~P.~Springer,
	%``${}^{3}\mathrm{He}$ and $pd$ Scattering to Next-to-Leading Order in Pionless Effective Field Theory,''
	Phys.\ Rev.\ C {\bf 89}, no. 6, 064003 (2014)
%	doi:10.1103/PhysRevC.89.064003
	[arXiv:1402.5441 [nucl-th]].
	%%CITATION = doi:10.1103/PhysRevC.89.064003;%%
	
	%\cite{Vanasse:2013sda}
	\bibitem{Vanasse:2013sda} 
	J.~Vanasse,
	%``Fully Perturbative Calculation of $nd$ Scattering to Next-to-next-to-leading-order,''
	Phys.\ Rev.\ C {\bf 88}, no. 4, 044001 (2013)
%	doi:10.1103/PhysRevC.88.044001
	[arXiv:1305.0283 [nucl-th]].
	%%CITATION = doi:10.1103/PhysRevC.88.044001;%%






\end{thebibliography}

\end{document}